\documentclass{aa}  

\usepackage{graphicx}

\usepackage{color}
\usepackage{ulem}
\usepackage{sidecap}
\usepackage{soul}
\usepackage{txfonts}
\begin{document}

    \title{Magnetic properties of the umbral boundary during sunspot decay}  
     \subtitle{Comparative study of multiple datasets}

  \titlerunning{Magnetic properties at the inner boundary of a decaying sunspot}

   \author{  M. Garc\'ia-Rivas  \inst{1,2},
             J. Jur\v c\'ak      \inst{1}
            \and
            N. Bello Gonz\'alez \inst{3}          
          }

   \institute{Astronomical Institute of the Czech Academy of Sciences, Fri\v cova 298, 25165 Ond\v rejov, Czech Republic \\
                \email{rivas@asu.cas.cz}    
                \and
                Astronomical Institute, Charles University, V Hole\v sovick\'ach 2, 18000 Praha, Czech Republic           
                \and
                Institut f\"ur Sonnenphysik (KIS), Georges-K\"ohler-Allee 401a, 79110 Freiburg, Germany
             }

   \date{Received 26 September 2023; accepted 14 June 2024 }

 
  \abstract
    {In recent years, the magnetic properties of the umbra-penumbra boundary of sunspots and the boundary of pores at various evolutionary stages have been characterised using datasets from different instruments. }
   {We aim to study the intrinsic differences between the intensity and vector magnetic field properties derived from Hinode/SP and SDO/HMI observations of a decaying sunspot.}
   {We analysed the sunspot embedded in active region NOAA 12797 during six days in  30 SP/Hinode scans and 704 HMI/SDO for both regular maps and maps corrected for scattered light, $\mathrm{HMI_{dcon}}$. We studied the correlation of the magnetic properties and continuum intensity in the datasets within the spot, and we investigated the differences at the umbra-penumbra boundary. We examined the decaying process in detail using the full temporal resolution of the $\mathrm{HMI_{dcon}}$ maps. }
   {We find a good one-to-one correspondence between the magnetic properties in the SP and $\mathrm{HMI_{dcon}}$ maps, but the continuum intensity of the spots in the SP maps is found to be $0.04\,I_\mathrm{QS}$ brighter than in the $\mathrm{HMI_{dcon}}$ maps. The considerable influence of scattered light in the HMI maps makes it the least ideal dataset for studying the boundary of spots without a penumbra. The properties at the umbra-penumbra boundary evolve slowly during the sunspot decay stage, while the penumbra still provides some stability. In contrast, they respond more abruptly to areal changes in the naked-spot stage. During the sunspot decay, we find linear decay in the area and in the magnetic flux. Moreover, the umbra shows two characteristic decaying processes: a slow decay during the first three days, and a sudden fast decay during the final dissipation of the penumbra. We find indications of a 3.5\,h lag between the dissipation of the vertical fields in the umbral region and the photometric decay of the umbral area.}
   {The differences found in the continuum intensity and in the vertical component of the magnetic field, $B_\mathrm{ver}$, between the analysed datasets explain the discrepancies among the $B_\mathrm{ver}$ values found at the boundaries of umbrae in previous studies.}


   \maketitle
   \nolinenumbers

\section{Introduction}

\begin{figure*}[!t]
\sidecaption
\includegraphics[width=1\linewidth]{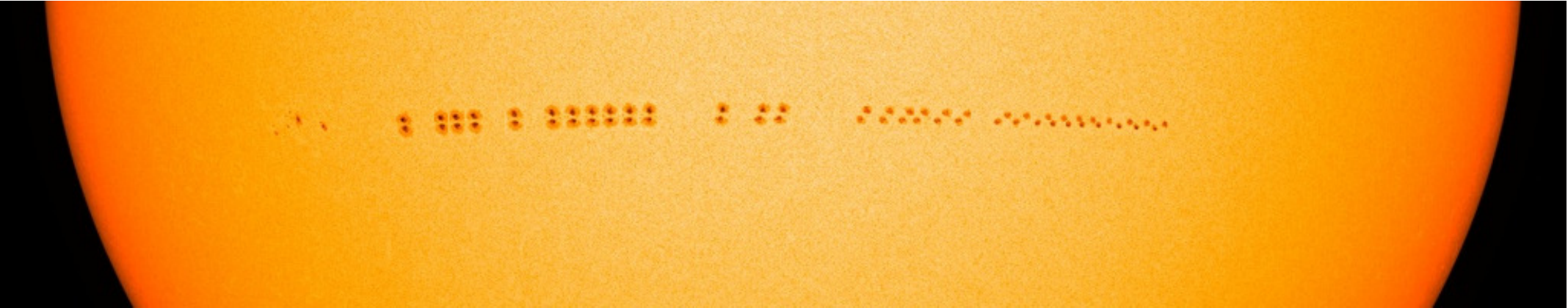}
\caption{HMI composition of the transition in AR NOAA~12797 from a decaying sunspot to naked spots. The 30 snapshots correspond to the scanning times of the analysed SP scans. AR NOAA~12798 on the left was magnetically connected to AR 12797 during most of the passage across the solar disc. }
\label{fig:evolution_AR}
\end{figure*}

Sunspots are the largest structures seen on the surface of the Sun. They are produced by the interaction of strong magnetism and convective motions. It is well known that sunspots consist of a dark inner part, the umbra, which is surrounded by lighter radial filaments, the penumbra. The umbra is produced by strong, vertical magnetic fields, while the penumbra is characterised by weaker and more horizontal magnetic fields. \cite{Chandrasekhar1961} studied the effects of magnetism on convection and stated that vertical fields ($B_\mathrm{ver}$) affect the convective mode, whereas horizontal fields shape the convective cells. 

Different methods have been used to characterise the boundary between umbrae and penumbrae (UP boundaries) in terms of a continuum intensity threshold, such as intensity distributions, cumulative histograms, or maximum intensity gradients \citep[e.g.][]{1981_Grossmann-Doerth,1997_Pettauer_cumulativeHist, 1993_Beck}. These methods have been applied, for example, to establish the relation between the sunspot area and brightness and the (in)variability of the area with the cycle phase \citep[e.g.][]{2007_Mathew}. Although this is useful for automatic sunspot detection analyses, each method has limitations. Many authors therefore rather select the intensity thresholds based on visual investigation and keep them constant during the evolution of the spot \citep[e.g.][]{2021Qialoning_alphas_decay, 2022_Li, Schmassmann_etal2018, Benko_etal2018}.

Sunspots are very dynamic structures. Despite this dynamism, \cite{Jurcak_etal2018} found a critical value of the vertical magnetic field ($B_\mathrm{ver}^\mathrm{crit} = 1867 \pm 18$\,G) at the UP boundaries of stable sunspots defined at 50\% of the continuum intensity of the local quiet Sun, $I_\mathrm{QS}$, based on a statistical study of maps from the Spectro-Polarimeter of the Solar Optical Telescope onboard the Hinode satellite (Hinode/SP). Further studies of stable sunspots and pores observed with different instruments have also found a $B_\mathrm{ver}^\mathrm{crit}$ value at the boundary \citep[e.g.,][found 1693~G in SDO/HMI\footnote{Helioseismic and Magnetic Imager on board the Solar Dynamics Observatory.} maps, 1787~G in GRIS\footnote{GREGOR Infrared Spectrograph.}/GREGOR maps, and 1731~G in SDO/$\mathrm{HMI_{dcon}}$\footnote{HMI maps corrected for scattered light}, respectively]{Schmassmann_etal2018, Lindner:2020, garciarivas_pore}. In contrast, \cite{Loptien2020_NoUniversalConnection} claimed that the invariable $B_\mathrm{ver}^\mathrm{crit}$ value is due to the use of a fixed continuum intensity threshold that falls at different positions of the UP boundary with sunspot sizes. That is, the intensity contour falls at different magnetic structures (spines and penumbral filaments) whose $B_\mathrm{ver}$ counteracts with each other and maintains a constant averaged vertical field.  

The question then is what happens when the magnetic structures undergo an unstable phase in their evolution. \cite{Jurcak_etal2015} investigated the process of the formation of a penumbra: The penumbra formed by invading the umbral regions with weak $B_\mathrm{ver}$ values until the incursion stabilised, leading to a UP boundary with a constant $B_\mathrm{ver}$ of the maximum observed value. During the formation of a penumbra around a pore of reduced $B_\mathrm{ver}$, the penumbra in contrast completely colonised the pore region, resulting in an orphan penumbra \citep{Jurcak_etal2017}. This line of investigation led to the Jur\v c\'ak criterion, an empirical law that states that UP boundaries of stable sunspots are not only defined by 50\% of the $I_\mathrm{QS}$, but also by a $B_\mathrm{ver}^\mathrm{crit}$ value. This criterion implies that umbral magnetoconvection settles in regions with $B_\mathrm{ver}~>~ B_\mathrm{ver}^\mathrm{crit}$. However, areas with $B_\mathrm{ver}~<~ B_\mathrm{ver}^\mathrm{crit}$ are unstable and prone to vanish against other more vigorous modes of (magneto\-)convection. The analysis of the Gough-Tayler stability criterion \citep{Gough:1966} in MURaM\footnote{Max-Planck-Institute for Solar System Research/University of Chicago Radiation Magneto-hydrodynamics.} sunspot simulations by \cite{Schmassmann_mhd} supports the importance of the role of $B_\mathrm{ver}$ in the stability of different modes of magneto-convection. 

The magnetic properties of a decaying sunspot observed by Hinode/SP were studied by \cite{Benko_etal2018}. They did not find a constant $B_\mathrm{ver}$ at the UP boundary, but rather a value that tended to weaken in the first stages and strengthened in the last stages of the decay. Moreover, the $B_\mathrm{ver}^\mathrm{crit}$ value found by \cite{Jurcak_etal2018} was only measured well inside the umbra during the seven-day process. In contrast, \cite{Loptien2020_NoUniversalConnection} presented a decaying sunspot with a constant $B_\mathrm{ver}$ at the UP boundary, equivalent to the value found by \cite{Jurcak_etal2018} during a five-day analysis of Hinode/SP maps. 

The physical processes leading to the disappearance of a penumbra are still not well understood. Moving magnetic features (MMFs) remove flux from the active region, but this is not always equivalent to the decay rate of the sunspot \cite[e.g.][]{2002MartinezPillet_decaySunspotMMF}. This means that other processes must play an important role as well, for instance, a change in the configuration of the magnetic canopy, similar to the process of the formation of the penumbra, but bottom-up instead \citep{2020Romano_restoringPenumbra}, or the erosion by convective motions \citep{2015Rempel_decaySunspot_nakedspot, Strecker+2021}. A difference in the  decay rates may explain why some decaying sunspots fulfil the Jur\v c\'ak criterion.

In this work, we analyse the properties on the UP boundary of a decaying sunspot that transforms into naked spots. We aim to study the differences with respect to other studies of decaying sunspots. We use $\mathrm{HMI_{dcon}}$ data, following the methods used in \cite{garciarivas_pore}. Complementary, since this spot was observed with Hinode/SP during its passage across the solar disc, we compare the magnetic properties derived from Hinode/SP, SDO/HMI and SDO/$\mathrm{HMI_{dcon}}$ to better understand the differences of  $B_\mathrm{ver}^\mathrm{crit}$ in the studies that were based on these datasets.


\section{Data}

In order to compare the magnetic properties derived from the Hinode/SP \citep{Hinode_Kosugi2007, SOT_Tsuneta2008} and SDO/HMI \citep{SDO_Pesnell2012, HMI_Schou2012} observations, we selected a long-lasting active region that traversed the solar disc and contained a decaying sunspot. The transition from sunspot to naked spots allows a comparison between the targets at different evolutionary stages.  

\subsection{Decaying sunspot}
 We studied a sunspot in active region (AR) NOAA~12797. This is an $\alpha$-active region with negative polarity, followed closely ($\sim100\arcsec$) by the $\beta$-active region NOAA~12798. The sunspot appeared on the eastern limb of the southern hemisphere on 17 January 2021. Initially, the umbra was divided by a light bridge and one penumbra surrounded both umbrae. During the following days, the penumbra around the light bridge vanished, while the two umbral cores grew apart. The complete dissipation of the penumbrae on 24 January led to two small naked spots. Instead of completely disappearing, they gathered more magnetic flux and grew larger before they completely disappeared from the solar surface close to the western limb on 28 January. The evolution of the spots is summarised in Fig.~\ref{fig:evolution_AR} with a composition of 30 HMI maps at the times of the analysed SP scans.

Despite having the spots on sight for several days, we limited the data to the period from 20 January at 18:36~UT to 26 January at 21:24~UT, which corresponds to the positions on the solar disc with $\mu > 0.75$, where $\mu\,=cos(\theta)$ and $\theta$ is the heliocentric angle. This constraint aimed to minimise the errors induced by the projection effects and to have a continuous dataset in HMI observations by avoiding data gaps. We also intended the overlap between the HMI and SP observations to be as long as possible. 

The first observation of a coronal mass ejection ever captured by the coronagraph Metis \citep{2020A_Antonucci_Metis} on board Solar Orbiter \citep{2020A_Muller_SolarOrbiter} occurred on 16 and 17 January 2021. The source region was declared consistent with the location of the active region of this study, AR~12797, on the eastern limb \citep{NOAA12797_Metis}. AR~12797 and mostly AR~12798 produced several B- and even C-class flares between the appearance on the eastern limb and the beginning of our study. Only weaker B-class flares continued to erupt in the neighbouring region AR~12798 from the beginning of our analysis until 23 January. Since the active regions were magnetically connected, these eruptions could have induced changes in the magnetic structure of the studied spots and, as a consequence, in the decay.

\subsection{SDO/HMI data}

In order to study the magnetic properties of the decaying sunspot, we required a homogeneous dataset with good temporal resolution, that is, a temporal cadence that allows for several observations during the fastest stages of the structural transitions. This requirement was largely fulfilled by HMI. It provides full-disc spectropolarimetric data every 12~minutes with a pixel scale of $\sim0,5\arcsec$. The filtergraph scans the \ion{Fe}{I} photospheric line at 617.3~nm along six positions with a spectral step of 7.6~pm. Ready-to-use data are provided by the Joint Science Operation Center (JSOC\footnote{\url{http://jsoc.stanford.edu/}}), where the standard full-disc vector magnetic field (\texttt{hmi.B\_720s}) and continuum intensity (\texttt{hmi.Ic\_720s}) maps can be downloaded. We required the best available spatial resolution to study the magnetic properties at the boundary of the spots. For this reason, we used the $\mathrm{HMI_{dcon}}$ series: regular HMI maps corrected for scattered light (\texttt{hmi.B\_720s\_dconS} and \texttt{hmi.Ic\_720s\_dconS}), as described in \cite{garciarivas_pore}. The $\mathrm{HMI_{dcon}}$ maps agree qualitatively with sub-arcsecond spatial resolution observations \citep[e.g.][]{HMI_dcon_Norton}, and they show a higher continuum intensity contrast in granulation and darker magnetic structures such as umbrae.

The two sets of Stokes parameters, standard HMI and $\mathrm{HMI_{dcon}}$, were processed similarly by the HMI vector magnetic field pipeline \citep{HMI_Pipeline_Hoeksema2014}: Full-disc photospheric vector magnetic field maps were inferred using the Milne-Eddington inversion code called Very Fast Inversion of the Stokes Vector \citep[VFISV,][]{VFISV_Borrero2011, VFISV_Centeno2014}. The Zeeman-induced 180-azimuth ambiguity was solved using the ME0 method, which is a variation of the minimum energy method \citep{disambiguation_metcalf,disambiguation_leka}, and the disambiguation solution was applied using \texttt{hmi\_disambig.pro}. We transformed the components of the magnetic field vector referred to the line of sight (LOS) (total magnetic field strength, LOS inclination, and azimuth) to components referred to a local reference frame (LRF) (vertical and transversal components) using the \texttt{hmi\_b2ptr.pro} routine. Following \cite{garciarivas_pore}, we analysed the total magnetic field strength (\textit{B}), the vertical magnetic field ($B_\mathrm{ver}$), and the LRF inclination ($\gamma = \mathrm{acos}{[B_\mathrm{ver}/B]}$) of 704 HMI and $\mathrm{HMI_{dcon}}$ cutouts.

Even though we analysed the sunspot for several days, the decaying process did not allow us to calculate the 24-hour orbital-induced magnetic variations reliably. Nonetheless, these variations were calculated by \cite{Schmassmann_etal2018} at the umbral boundary of a stable sunspot during $\sim$\,10 days, where $B$ and $B_\mathrm{ver}$ oscillated by less than 20~G and $\gamma$ oscillated by $0.2\,^{\circ}$. Therefore, we can neglect the 24-hour oscillations as their magnitude is significantly lower than the actual change in the magnetic parameters caused by the sunspot decay.

\subsection{Hinode/SP data}
The SOT incorporates the SP \citep{Ichimoto2008}, which is a slit-spectropolarimeter that provides high-resolution solar observations with spatial samplings along the slit and in the scanning direction, and temporal cadences that depend on the mapping mode. The current dataset was recorded with a spatial scale of $0.30\arcsec\times 0.32\arcsec$ and a temporal cadence of 1.6~s per slit position, that is, in fast mode. The SP spectral range is centred on two Zeeman-sensitive Fe~I photospheric lines (at 630.15 and 630.25~nm), including the nearby continuum, and has a spectral sampling of 2.15~pm. The different magnetic sensitivities of the spectral lines permit a better accuracy in the determination of the vector magnetic field. Even though different types of data are publicly available (e.g. \texttt{Level 0} raw non-calibrated Stokes profiles from one slit position; \texttt{Level 1} calibrated Stokes profiles from a full scan), we used already inverted data (\texttt{Level 2} and \texttt{2.1}) provided by the Community Spectro-polarimetric Analysis Center (CSAC\footnote{\url{https://csac.hao.ucar.edu/sp_data.php}}).

The inversion, that is, inferring the magnetic and thermodynamical properties at the line formation height, was made with the Milne-Eddington gRid Linear Inversion Network \citep[MERLIN,][]{2007_Lites_MERLIN}. Among other parameters outside of our current interest, Level 2 data provide the vector magnetic field with respect to the LOS reference frame (magnetic field flux ($B_f$), LOS inclination, non-disambiguated azimuth, and filling factor ($\alpha$)). \texttt{Level 2.1} data contain the vector magnetic field (vertical and transversal) with respect to the LRF and the disambiguated azimuth used to transform the reference frame. The disambiguation was carried out using the same algorithm as was used in HMI data, ME0. In contrast to HMI inversions, where the filling factor is constant ($\alpha$=1), $\alpha$ is variable in SP inversions. Therefore, the total magnetic field strength was calculated by multiplying the inverted field flux by the filling factor ($B = \alpha B_f$). The LRF inclination was obtained from the ratio of the total magnetic field strengths and vertical fields as for HMI datasets. 

As a result, a combination of \texttt{Level 2} and \texttt{Level 2.1} data allows us to retrieve the same parameters as were analysed from HMI (\textit{B}, $B_\mathrm{ver}$, and $\gamma$) from 30 scans. Six observations were removed from the dataset due to interruptions in the scanning. Since a full scan takes approximately 45~min and scans are not taken consecutively, the temporal resolution is poor for our purposes. However, the high spatial resolution allows a more precise analysis of the UP magnetic properties than HMI maps. 

This is the first analysis that uses \texttt{Level 2} and \texttt{Level 2.1} SP data to assess the $B_\mathrm{ver}$ evolution. Previous analyses, namely \citet{Jurcak_etal2018,Benko_etal2018}, used the Stokes Inversion based on Response functions code \citep[SIR, ][]{SIR_RuizCobo1992} with constant magnetic parameters in the modelled atmosphere. This inversion scheme of the SIR code, as well as any inversion code based on the Milne-Eddington approximation, provides us with an average value of the magnetic field parameters at optical depths where the line is most sensitive to these parameters. According to \citet{CabreraSolana:2005}, the spectral lines observed by SP and HMI are most sensitive to magnetic field parameters between $\log \tau_\mathrm{c} = -1$ and $-2$. The results from a similar range of optical depths were used by \citet{Jurcak2011}, who used the SIR code with height-dependent stratifications of the magnetic parameters. 

With the decay of the sunspot, we expect changes in the Wilson depression \citep{Loptien:2018}, and thus, the geometrical height of the region in which we obtain the magnetic parameters increases during the decaying process. As summarised by \citet{Balthasar:2018}, the magnetic field strength and also its vertical component decrease with height in the atmosphere. Combining the results of \citet{Loptien:2018} and \citet{Balthasar:2018}, we estimate that $B$ and $B_\mathrm{ver}$ decrease up to 100~G due to the increase in the geometrical height during the decay process. However, we cannot determine this value exactly with the applied data analysis methods.

The methods for comparing the magnetic properties derived from the HMI, $\mathrm{HMI_{dcon}}$, and SP maps at the decaying spot and at the umbral boundary are described in Sect.~\ref{sec:methods}. The results are summarised in Sect.~\ref{sec:results}, and a discussion is provided in Sect.~\ref{sec:conclusions}.

\section{Analysis of the method}\label{sec:methods}

Since the aim of our current research is to study the evolution of the magnetic properties at the boundary of dark structures (sunspots or pores), and because different studies have been carried out using a variety of instruments \citep[e.g., Hinode/SP, SDO/HMI, SDO/$\mathrm{HMI_{dcon}}$ by][]{Jurcak_etal2018, Schmassmann_etal2018, Benko_etal2018, 2023_Poros_JoseIvan, garciarivas_pore}, we find it necessary to study the differences between these datasets in more detail. We only investigated the magnetic parameters in spots areas and disregarded the rest of the field of view (FOV). In the case of a decaying sunspot, the first step is to examine side by side the magnetic properties of the whole spot from the SP, HMI, and $\mathrm{HMI_{dcon}}$ observations. To do this, the outer boundary has to be defined. Moreover, the umbra-penumbra (UP) boundary has to be characterised with the aim of studying the temporal evolution of the magnetic properties in it and comparing the different datasets. These two boundaries are defined in terms of continuum intensity thresholds.

\begin{table}[!t]
\caption{Summary of the SP scans. }              
\label{table:observations}      
\centering                          
\begin{tabular}{c c c c c}       
\hline\hline      
No. & Time (UT)   & Time (UT)          & Time (UT)  & $\mu$ \\
        & SP scan & SP slit on spots & HMI map & \\
\hline     \hline     
      & & & & \\
    1 & 21 Jan 06:41:28 &   07:04:00    & 07:00:00   & 0.83 \\      
    2 & 21 Jan 13:00:03 &  13:23:00     & 13:24:00   & 0.86 \\
    3 & 21 Jan 15:35:04 &   15:58:00    & 16:00:00   & 0.87 \\
    4 & 21 Jan 18:36:28 &   18:59:00    & 19:00:00   & 0.88 \\
    5 & 22 Jan 01:00:04 &    01:23:00   & 01:24:00   & 0.90 \\   
    6 & 22 Jan 07:05:05 &   07:28:00    &  07:24:00  & 0.92 \\
    7 & 22 Jan 10:20:03 &   10:43:00    & 10:48:00   & 0.93 \\
    8 & 22 Jan 13:30:03 &  13:53:00     & 13:48:00   & 0.94 \\
    9 & 22 Jan 16:10:03  &   16:33:00   &  16:36:00  & 0.95 \\
    10 & 22 Jan 19:11:27  &  19:35:00    & 19:36:00  & 0.95 \\      
    11 & 22 Jan 22:13:03 &   22:36:00   &  22:36:00  & 0.96 \\
    12 & 23 Jan 09:17:03 &   09:40:00   & 09:36:00   & 0.97 \\
    13 & 23 Jan 15:11:33 &   15:35:00   & 15:36:00   & 0.97 \\
    14 & 23 Jan 18:14:26 &    18:37:00   & 18:36:00  & 0.97  \\      
    15 & 24 Jan 06:41:26 &   07:03:00    & 07:00:00  & 0.97 \\
    16 & 24 Jan 10:00:02 &   10:22:00    & 10:24:00  & 0.96 \\
    17 & 24 Jan 13:00:02 &   13:22:00   &  13:24:00  & 0.96 \\
    18 & 24 Jan 15:30:02 &    15:51:00  & 15:48:00   & 0.96  \\
    19 & 24 Jan 18:35:01 &  18:56:00    & 19:00:00   & 0.95  \\      
    20 & 24 Jan 21:50:01 &    22:11:00  &  22:12:00  & 0.94 \\ 
    21 & 25 Jan 04:00:01 &   04:20:00   & 04:24:00   & 0.93 \\
    22 & 25 Jan 07:16:25 &   07:37:00   & 07:36:00   & 0.92 \\
    23 & 25 Jan 10:34:01  &   10:55:00   & 11:00:00  & 0.91 \\     
    24 & 25 Jan 13:15:01  &  13:35:00    & 13:36:00  & 0.90  \\
    25 & 25 Jan 15:45:01  &   16:05:00   & 16:00:00  & 0.89  \\
    26 & 25 Jan 17:50:01 & 18:10:00     & 18:12:00   & 0.88 \\
    27 & 25 Jan 20:45:01  &  21:05:00    & 21:00:00  & 0.87 \\
    28 & 26 Jan 00:00:01  &  00:20:00    & 00:24:00  & 0.86  \\      
    29 & 26 Jan 03:00:01 &  03:20:00    & 03:24:00   & 0.84  \\
    30 & 26 Jan 06:33:25 &  06:53:00    & 06:48:00   & 0.83 \\
   
\end{tabular}
\tablefoot{Second column: SP scan time, corresponding to the beginning of the scan. Third column: Time when the SP slit observed the middle of the spots. Fourth column: Time of the HMI maps closest to the SP observations in the middle of the spots.}
\end{table}

As described above, the temporal and spatial sampling of SP and HMI is not identical. In order to directly compare the magnetic properties derived from the two instruments, the data with the best pixel scale (SP, $\sim0.30\arcsec\times0.32\arcsec$) must be degraded to match the lowest spatial sampling (HMI, $0.5\arcsec\times0.5\arcsec$). SP maps were aligned and resampled to HMI maps using \texttt{auto\_align\_images.pro}, a routine that minimises the cross-correlation between images while allowing for image warping, if needed. The warping is allowed as a way to remove possible effects of irregular slit movements in SP scans  \citep{2009_Centeno_noregslitscan_SP}. First, reference points in the two images provide rough transformation parameters via \texttt{caltrans.pro}. These serve as initial transformation parameters for a first run of \texttt{auto\_align\_images.pro}, using the minimisation algorithm Amoeba. The obtained transformation parameters are used as the initial transformation parameters of the final run of \texttt{auto\_align\_images.pro}, now allowing for image warping and allowing the more robust minimisation algorithm Powell. The result is PQ parameters (shifts, scales, rotations, etc.) that transform each SP scan into a reference HMI map. Since SP scans are a combination of several slit positions at different times, we must find the time at which the slit was in the middle of the spots, and align this to the HMI map closer to that time (see table \ref{table:observations}). On average, SP spends around 5~min to cover the whole spot area, while HMI data are a sum of 12 min full-disc observations. Therefore, we expect evolutionary differences in the compared maps that affect the alignment procedure and our results, but these cannot be removed. As a way to minimise the effects of granular evolution in the alignment procedure, the transformation was carried out by masking pixels with a continuum intensity brighter than 95\,\% of the continuum intensity of the local quiet Sun ($I_\mathrm{c}$>0.95\,$I_\mathrm{QS}$). 

The PQ parameters are not only a tool for transforming SP scans into HMI maps, but also for transforming the boundary contours from HMI and $\mathrm{HMI_{dcon}}$ into SP maps and vice versa by means of \texttt{pq2xy.pro} and \texttt{pqinvert.pro}. This approach permits us to compare the one-to-one correspondence between magnetic and intensity maps and to establish a relation between the UP boundary properties in the different non-transformed datasets.

\begin{figure}[!t]
\includegraphics[width=1\linewidth]{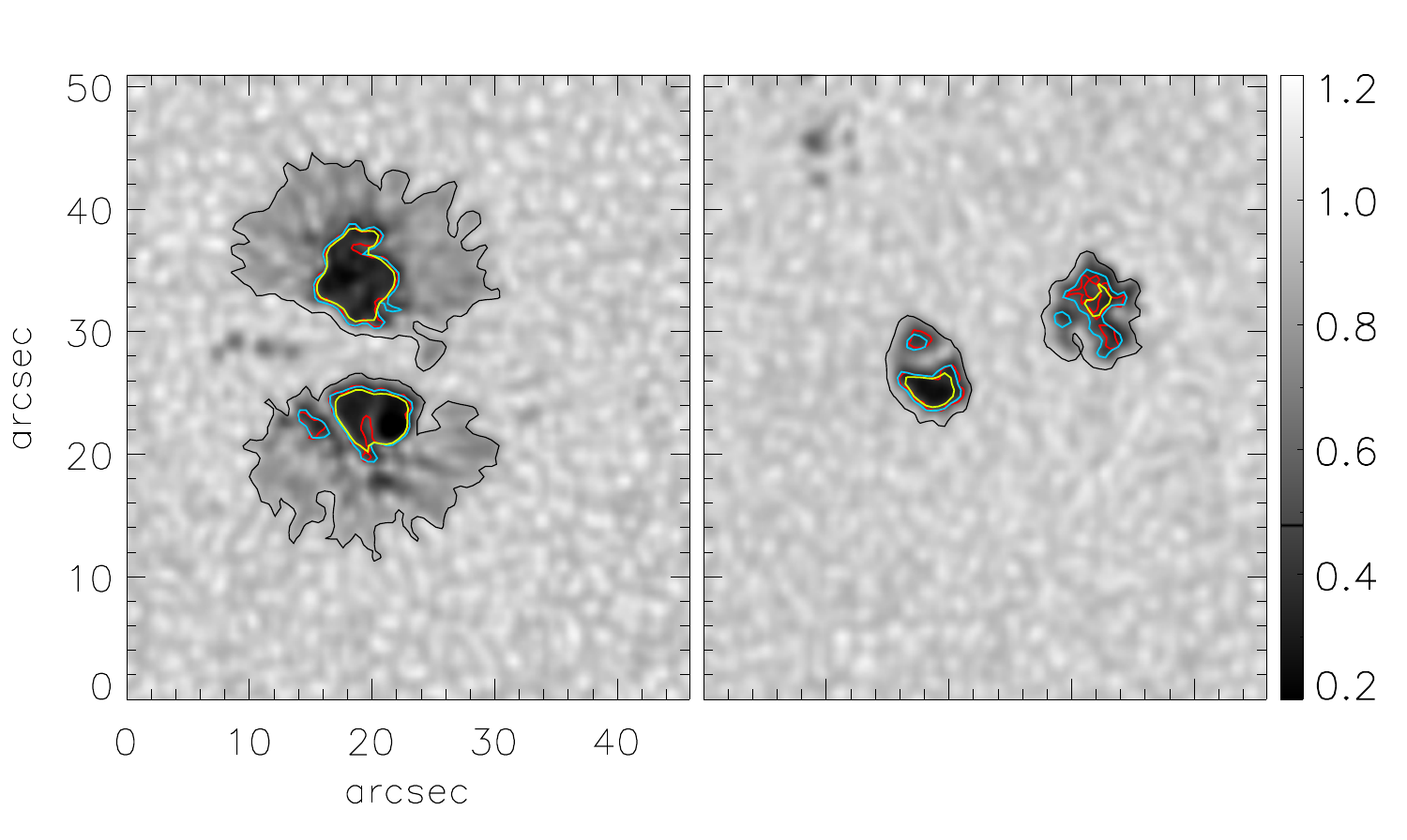}
\caption{Examples of continuum intensity $\mathrm{HMI_{dcon}}$ maps. The black lines outline the $I_\mathrm{c}=0.90I_\mathrm{QS}$ threshold. The colour lines mark the $I_\mathrm{c}=0.50I_\mathrm{QS}$ location in the SP (red), HMI (yellow) and $\mathrm{HMI_{dcon}}$ (blue) co-temporal maps. Left: Sunspot stage on 22 January at 10:48~UT. Right: Naked-spots stage on 26 January at 06:48~UT.}
\label{fig:maps_all_contours}
\end{figure}

\begin{figure*}[!t]
 \includegraphics[width=\linewidth]{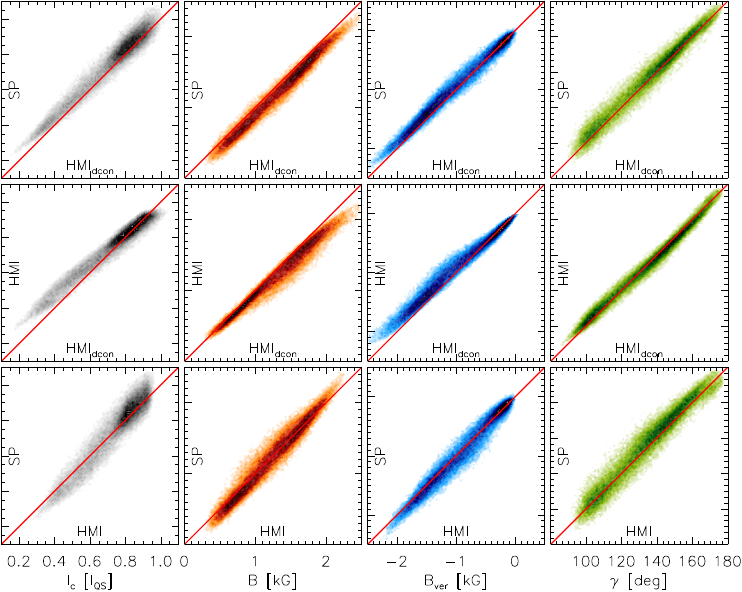}
 \caption{Density functions comparing the continuum intensity ($I_\mathrm{c}$, grey), total magnetic field ($B$, orange), vertical magnetic field ($B_\mathrm{ver}$, blue), and magnetic field inclination ($\gamma$, green) between the SP, HMI, and $\mathrm{HMI_{dcon}}$ datasets. Top row: SP vs. $\mathrm{HMI_{dcon}}$. Middle row: HMI vs.  $\mathrm{HMI_{dcon}}$. Bottom row: SP vs. HMI. The red line marks the one-to-one correspondence. The spans of the $y$-axes are identical to those of the $x$-axes.} 
\label{fig:scatter_plots}
\end{figure*}

\begin{figure*}[!t]

\hfill\includegraphics[width=0.8\linewidth]{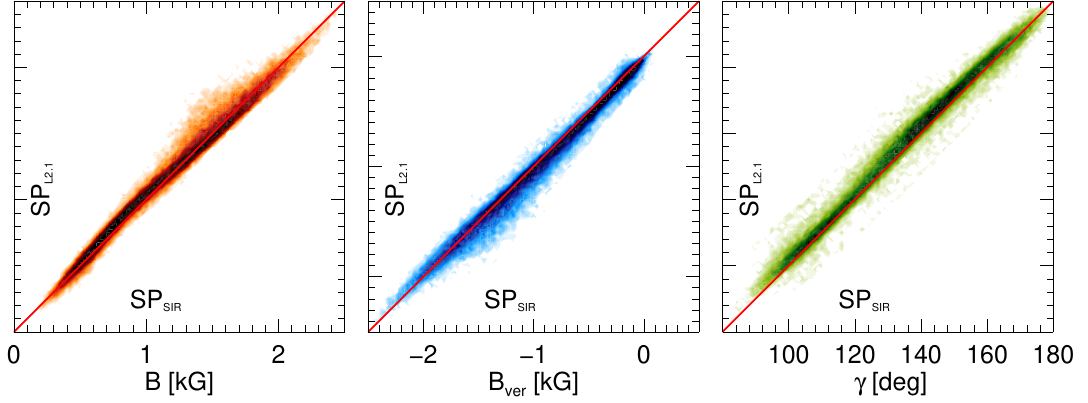}

\caption{SP density functions comparing the total magnetic field ($B$, orange), vertical magnetic field ($B_\mathrm{ver}$, blue), and inclination ($\gamma$, green) inverted with SIR ($_\mathrm{SIR}$) and MERLIN ($_\mathrm{L2.1}$). }
\label{fig:SP_v_JJ}
 
\end{figure*}

The continuum intensity thresholds were delimited from $\mathrm{HMI_{dcon}}$ maps. The $\mathrm{HMI_{dcon}}$ data suit our goals better because they provide a very good temporal cadence and improved sharpness. We therefore used them as the reference maps. The external boundary of the spots was defined at $I_\mathrm{c}=0.90I_\mathrm{QS}$ based on visual inspection. In the early stages of the decay, this threshold traces the penumbral outer boundary, while in the later stages, it outlines weakly magnetised regions of the naked spots. Following the method used by \cite{Jurcak_etal2018}, the internal boundary was defined at $I_\mathrm{c}=0.50I_\mathrm{QS}$, which in the early stages delimits the UP boundary, and in the later stages, it outlines the darker umbral cores. We recall that the sunspot occasionally contains small darker regions that are embedded in the penumbra, while the naked spots are formed by various darker umbral cores that are surrounded by dark areas that are slightly brighter. Only the larger dark regions were considered in the investigation; contours that encircled regions smaller than $0.75$~Mm$^2$ were neglected. Figure~\ref{fig:maps_all_contours} shows two stages of the transition, the sunspot on the left and naked spots on the right, with the mentioned contours. The better spatial resolution of SP is perceived in the more corrugated boundary contours that show intrusions into the umbral region of the sunspot and a division into multiple darker cores in the naked spots. The discarded small umbral cores are visible as the regions with an $\mathrm{HMI_{dcon}}$ blue contour without an SP red contour. On the other hand, the yellow HMI contours fail to reach the full extent of the spots and the corrugation of the UP boundary. To conclude, the temporal evolution of the magnetic properties at the internal boundaries was studied by averaging $B$, $B_\mathrm{ver}$, and $\gamma$ along the $I_\mathrm{c}=0.50I_\mathrm{QS}$ threshold based on $\mathrm{HMI_{dcon}}$ intensities.

\section{Results}\label{sec:results}


\subsection{Comparison between datasets within the spot}\label{subsec:scatters}

To better understand the differences between SP, HMI, and $\mathrm{HMI_{dcon}}$, a one-to-one comparison is required. Only pixels within the spot (90\% of $I_\mathrm{QS}$ in $\mathrm{HMI_{dcon}}$ maps, or from now on, $I_\mathrm{c,dcon}=0.90I_\mathrm{QS}$) were considered. Figure~\ref{fig:scatter_plots} displays the two-dimensional density functions of the compared datasets. Additionally, Fig. \ref{fig:SP_v_JJ} compares the SP vector fields inferred with the MERLIN (the default dataset used in this work) and SIR codes. Unless specified otherwise, any reference to SP data refers to the vector field retrieved from \texttt{Level 2} and \texttt{Level 2.1} datasets, which were inferred with MERLIN.    

For SP versus $\mathrm{HMI_{dcon}}$ (top row in Fig.~\ref{fig:scatter_plots}), the continuum intensity is $\sim5~\%$ darker in $\mathrm{HMI_{dcon}}$ maps than in SP on average. The one-to-one correspondence is better for the total magnetic field strength, but there is a $\sim$100\,G offset where $\mathrm{HMI_{dcon}}$ values are stronger than those in SP. On the other hand, $B_\mathrm{ver}$ values are comparable at the lowest values of the magnetic field, but the $\mathrm{HMI_{dcon}}$ values we retrieved are stronger on average, with a $\sim$70\,G offset with respect to SP. The inclination, $\gamma$, has a linear trend in the whole range and an offset of $\sim$\,(1-2)$^\circ$ in the umbral region. Penumbral regions are retrieved to be more vertical in SP, and the scattering is stronger than in umbral regions.

For HMI versus $\mathrm{HMI_{dcon}}$ (middle row in Fig.~\ref{fig:scatter_plots}), we limited the 704 available maps to the 30 maps used for SP in order to obtain similar statistics. The continuum intensity does not have a one-to-one correlation; while bright areas are darker in HMI than in $\mathrm{HMI_{dcon}}$, the dark areas are brighter in HMI than in $\mathrm{HMI_{dcon}}$. $B$ and $B_\mathrm{ver}$ are similar in their magnetic properties in HMI compared to $\mathrm{HMI_{dcon}}$ for the weaker values, and the differences increase linearly with the parameter amplitudes. The inclinations are comparable in the whole range of obtained values.

\begin{figure*}[!t]
\includegraphics[width=\linewidth]{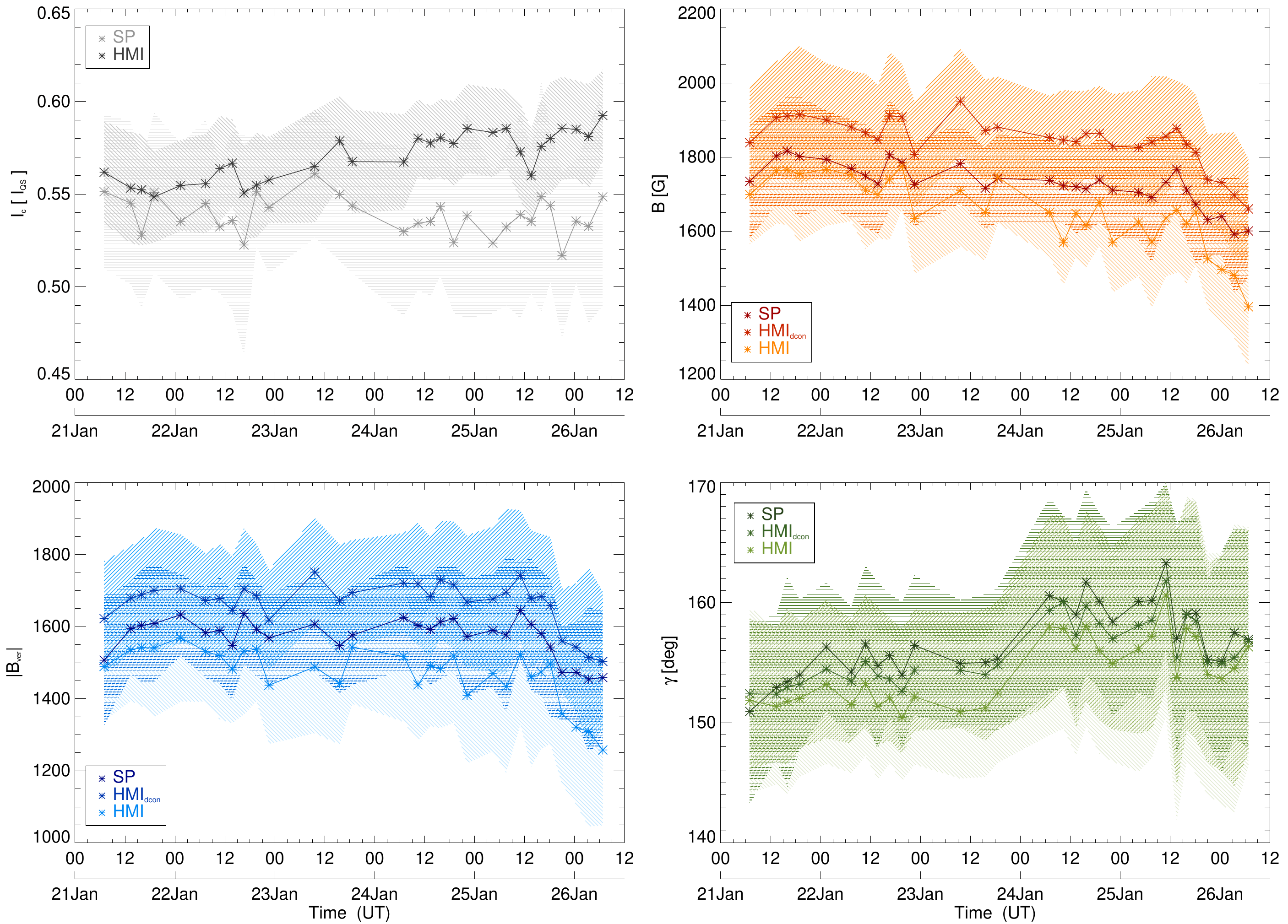}
\caption{Temporal evolution of the properties averaged along the $\mathrm{HMI_{dcon}}$ threshold $I_\mathrm{c}=0.50\,I_\mathrm{QS}$ in the SP, HMI, and $\mathrm{HMI_{dcon}}$ maps. The striped areas illustrate the standard deviation of the averaged properties ($\beta=0^\circ$: SP, \textit{$\beta=45^\circ$}: $\mathrm{HMI_{dcon}}$, and \textit{$\beta=135^\circ$}: HMI). Top left: Continuum intensity ($I_\mathrm{qs}$). Top right: Total magnetic field strength ($B$). Bottom left: Vertical magnetic field ($B_\mathrm{ver}$). Bottom right: LRF inclination ($\gamma$). }
\label{fig:050HMI_inALL}
\end{figure*}

We also compared SP and HMI (bottom row in Fig.~\ref{fig:scatter_plots}) because previous studies related to the magnetic nature of UP boundaries have used both datasets. The SP versus HMI density plots exhibit the greatest dispersion in all the parameters. Regarding $I_\mathrm{c}$, bright pixels are brighter in SP than in HMI, while dark pixels are darker in SP maps. $B$ is retrieved to be weaker in SP maps, where the field is lower, and it is stronger in the strongest fields. Similarly, $B_\mathrm{ver}$ is retrieved to be stronger in SP than in HMI in the regions with stronger fields, but the weaker fields are retrieved to be similar. The inclination has a good one-to-one correspondence in the more horizontal fields and an offset of $\sim$\,3$^{\circ}$ in the umbral region, where the SP values are more vertical than the HMI values.

We note that the $B_\mathrm{ver}$ parameter has the best correlation of all the datasets and the best one-to-one correspondence between SP and $\mathrm{HMI_{dcon}}$. Complementary, $\gamma$ has the best one-to-one correspondence between SP and HMI maps and between HMI and $\mathrm{HMI_{dcon}}$ maps. On the other hand, the $I_\mathrm{c}$ maps have the worst one-to-one correlation of all the parameters in all the compared datasets. Therefore, a magnetic field threshold, that is, a fixed $B_\mathrm{ver}$ value, would outline more similar umbral regions between different datasets than a fixed $I_\mathrm{c}$ threshold.

We additionally compared the default dataset analysed in this work (the vector field components inferred with MERLIN) to the vector field obtained with SIR. Figure \ref{fig:SP_v_JJ} shows a relatively good correspondence in the $B$, $B_\mathrm{ver}$, and $\gamma$ maps for SIR and MERLIN datasets. Based on the regions with the highest density function, MERLIN retrieves $\sim$\,40\,G higher values of $B$ and $B_\mathrm{ver}$ in weaker magnetic fields (<1.4\,kG), while SIR retrieves higher values of $B$ ($\sim$~30~G) in stronger magnetic fields (>1.4\,kG). Strong $B_\mathrm{ver}$ fields have a good one-to-one correspondence. A region with stronger scattering in $B$ ([1.2 - 1.8]\,kG) and in $B_\mathrm{ver}$ ([-1.0,-1.6]\,kG) corresponds to the naked-spot stage, which shows that during the absence of a penumbra, the MERLIN magnetic fields can be retrieved to be up to $\sim$150\,G stronger than SIR magnetic fields.  $\gamma$ is $1-2^{\circ}$ more vertical in vector fields obtained with MERLIN than with SIR on average, independently of the spot stage. 

\subsection{Comparison between datasets at the UP boundary}

After the general comparison of the spot properties inferred from interpolated SP, HMI, and $\mathrm{HMI_{dcon}}$ maps, we investigated the differences of the physical parameters among different datasets specifically at the UP boundary. 

As already mentioned, $\mathrm{HMI_{dcon}}$ is the main dataset of interest in this study. Therefore, we used the $I_\mathrm{c}=0.50I_\mathrm{QS}$ threshold defined in $\mathrm{HMI_{dcon}}$ maps ($I_\mathrm{c,dcon}=0.50I_\mathrm{QS}$) as a reference and projected it onto SP and HMI maps. Since HMI and $\mathrm{HMI_{dcon}}$ have the same pixel scale, the boundary coordinates are the same. On the other hand, the boundary coordinates in the original SP maps are obtained thanks to the PQ transformation parameters. We emphasize that the UP boundaries were studied on the original maps, not in transformed maps, as was done in subsection \ref{subsec:scatters}. The evolution of the physical properties averaged along the $I_\mathrm{c,dcon}=0.50I_\mathrm{QS}$ contour is shown in Fig.~\ref{fig:050HMI_inALL}.

In the case of $I_\mathrm{c}$ (upper left, grey in Fig.~\ref{fig:050HMI_inALL}), the UP boundary crosses brighter regions in SP and HMI maps on average. In order to address the significance of the magnetic structure (spot or naked spot), we compared $I_\mathrm{c}$ averaged during the first day (21 January; sunspots with a developed penumbrae) and last day (26 January; naked spots). In the SP maps, the UP boundary crosses structures with an averaged intensity of $I_\mathrm{c}=0.54I_\mathrm{QS}$ independently of the observed structure. Even though this result was expected from Fig.~\ref{fig:scatter_plots}, the significant standard deviation in the SP maps suggests that another factor affects the brightness at the contour. The larger $\mathrm{HMI_{dcon}}$ pixel scale translates into maps with fewer small-scale details. Small light bridges, for instance, are not well resolved. However, as shown in Fig.~\ref{fig:maps_all_contours}, SP contours are more intricate, and as a result, $I_\mathrm{c,dcon}=0.50I_\mathrm{QS}$ crosses the SP UP boundary, the brighter light-bridge ends, and even penumbral filaments, increasing both the averaged $I_\mathrm{c}$ and the standard deviation values. 

In the HMI maps, the average $I_\mathrm{c}$ at the UP boundary is $I_\mathrm{c}=0.57I_\mathrm{QS}$. The tendency of increasing $I_\mathrm{c}$ with time is clear in this case. The last-day observations are 3\% brighter than the first-day observations, that is, $I_\mathrm{c}=0.58I_\mathrm{QS}$ on the last day compared to $I_\mathrm{c}=0.55I_\mathrm{QS}$ on the first day. The scattered light from the neighbouring granulation in naked spots or pores affects the characterisation of the spot boundary in terms of a $I_\mathrm{c}$ threshold up to $\sim$3\% compared to the phase when the UP boundary is influenced by the scattered light from a darker penumbra.

In the case of the total field strength, $B$ (upper right, orange in Fig.~\ref{fig:050HMI_inALL}), three temporal evolutions are depicted (for SP, $\mathrm{HMI_{dcon}}$, and HMI data). Overall, a decrease in the total magnetic field strength is observed in all the datasets due to the decaying process. The total magnetic field is derived to be weaker by $\sim\,6\%$ in the SP than in the $\mathrm{HMI_{dcon}}$ datasets during the whole period on average, even though their difference varies by $\sim\,2\%$ with the position of the spot on the solar disc: the larger $\mu$, the smaller the difference between datasets. On the other hand, the evolutionary state of the spot causes no difference. 

As for $B$ in HMI maps, the evolution of the divergence is more obvious. The effects of $\mu$ are not evident since both types of observations are processed in the exact same way. We only see a slowly increasing divergence between datasets. HMI is weaker by 10\% than $\mathrm{HMI_{dcon}}$ on average, starting at 8\% weaker during 21 January, and increasing up to 16\% at the end of the observations. This increase is related to the size of the structure, and thus, to the increasing influence of the scattered light towards the end of the analysed period.

The magnetic field inclination, $\gamma$ (lower right, green in Fig.~\ref{fig:050HMI_inALL}) increases (it becomes more vertical with respect to the solar surface) during the first three days, followed by a decrease (it becomes more horizontal) in the later stages of the naked spots. The inclination is inferred to be more vertical in SP maps than in the two HMI maps, except for the first observation. The inclination in the SP maps is lower than 1\,\%  and more vertical than in $\mathrm{HMI_{dcon}}$ maps, while the inclination in HMI maps is 1\,\% more horizontal than in $\mathrm{HMI_{dcon}}$ maps on average. The difference between HMI and $\mathrm{HMI_{dcon}}$ varies with $\mu$, and it is smaller than 1\,\% close to the limb and around 2\,\% close to disc centre. Fig.~\ref{fig:scatter_plots} also shows that the retrieved inclinations are comparable in all the datasets.

\begin{figure*}[t!]
  \begin{minipage}[c]{0.75\linewidth}
    \includegraphics[width=0.95\linewidth]{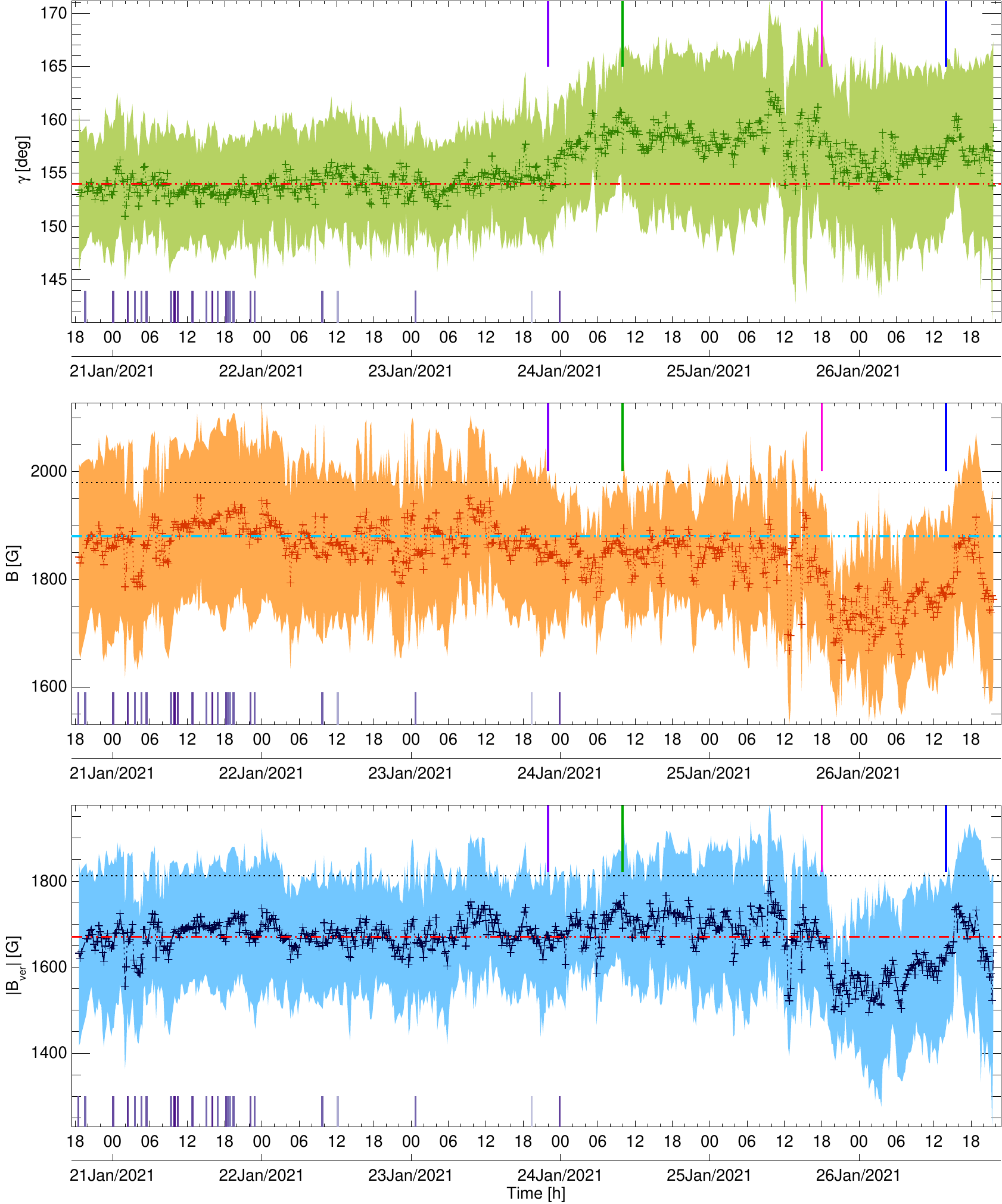}
  \end{minipage}
  \begin{minipage}[c]{0.25\linewidth}
  \vspace{6cm}
    \caption{Temporal evolution of the magnetic properties: $\gamma$ (in green, top), $B$ (in orange, middle), and $B_\mathrm{ver}$ (in blue, bottom) averaged along the UP boundary in $\mathrm{HMI_{dcon}}$  maps. The filled areas stand for the standard deviation of the averaged properties. The dotted horizontal lines (1821~G and 1979~G) mark the maximum \textit{B} and \textit{$B_\mathrm{ver}$} found in an evolving pore in $\mathrm{HMI_{dcon}}$ maps in \cite{garciarivas_pore}, but retrieved from the $I_\mathrm{c,dcon}=0.50I_\mathrm{QS}$ contour instead of the $I_\mathrm{c,dcon}=0.55I_\mathrm{QS}$ reported in the study. The upper vertical lines indicate the times of specific evolutionary stages: The beginning of the fast decay (purple), the end of the fast decay (green), the accumulation of new flux (pink), and the decrease in the spot area (blue). The bottom vertical lines indicate the flaring times of AR 12798. The colour relates to the GOES flare class, from lighter purple (weakest: A5.6) to darker purple (strongest: B5.4).}
    \label{fig:magnetic_evolution}
  \end{minipage}
\end{figure*}

\begin{figure*}[t!]
  \begin{minipage}[c]{0.8\linewidth}
    \includegraphics[width=\linewidth]{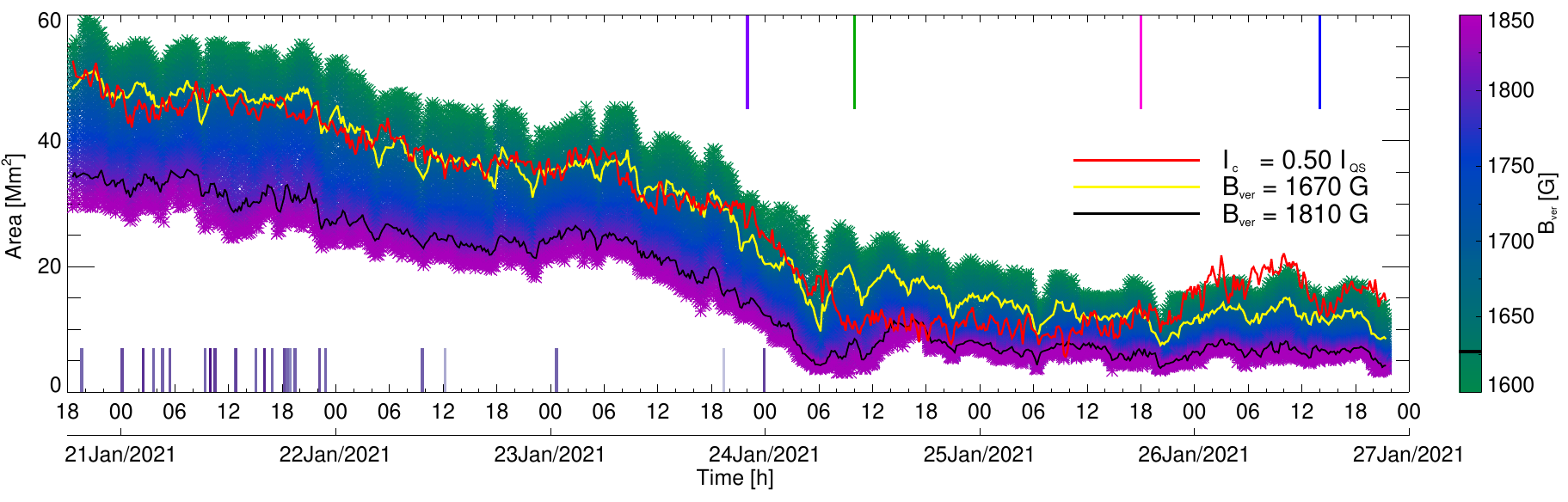}
  \end{minipage}
  \begin{minipage}[c]{0.19\linewidth}
   \caption{Temporal evolution of the areas encircled by $B_\mathrm{ver}$ thresholds (1600-1850~G) compared to the areas encircled by the threshold $I_\mathrm{c,dcon}=0.50I_\mathrm{QS}$ (yellow). The contours $B_\mathrm{ver}=1810$~G and $B_\mathrm{ver}=1670$~G are highlighted (black and yellow, respectively). The upper and lower vertical lines are the same as in Fig. \ref{fig:magnetic_evolution}.}
    \label{fig:all_bver}
  \end{minipage}
\end{figure*}

The vertical magnetic field (lower left, blue in Fig.~\ref{fig:050HMI_inALL}) shows a combination of the results from $B$ and $\gamma$. During most of the observations, $B_\mathrm{ver}$ oscillates around a fixed value in SP and $\mathrm{HMI_{dcon}}$ during the decay of the penumbra, and it decreases during the last stages of the naked spots, which correspond to the emergence of new flux, as discussed in section \ref{subsec:tev_hmi}. The values of $B_\mathrm{ver}$ in $\mathrm{HMI_{dcon}}$ maps are stronger by 6\% on average than in SP maps. The difference between the two datasets varies slightly with $\mu$. It reaches a minimum difference at the western limb ($\sim$\,4\%) and a maximum difference close to the disc centre ($\sim$\,6\%).

$B_\mathrm{ver}$ in HMI weakens constantly, with fluctuations, during the whole observation period. The difference between HMI and $\mathrm{HMI_{dcon}}$ $B_\mathrm{ver}$ values mostly depends on the spot size: On 21 January, $B_\mathrm{ver}$ in HMI is 9\,\% weaker than in $\mathrm{HMI_{dcon}}$, and it continuously decreases by 15\,\% on 26 January.

\subsection{Evolution of the decaying sunspot in $\mathrm{HMI_{dcon}}$} \label{subsec:tev_hmi}

In this section, we focus on the detailed study of the transition from a decaying sunspot to naked spots. This case study is interesting enough on its own, and it allows us to compare it to other studies of decaying sunspots. Since we aim to investigate the magnetic properties as accurately as possible, we used $\mathrm{HMI_{dcon}}$ maps with the full 12 min temporal cadence (704 maps from 20 January at 18:36~UT to 26 January at 21:24~UT). The magnetic properties averaged along $I_\mathrm{c,dcon}=0.50I_\mathrm{QS}$ are shown in Fig.~\ref{fig:magnetic_evolution}. We included the flaring times of the neighbouring region (AR 12798), to which the analysed AR was magnetically connected. Due to the weak nature of the flares, we did not observe any clear change due to the flaring activity. Additionally, we marked interesting times of the decay, such as the beginning and end of the fast decay during the last stages of the sunspot, the accumulation of magnetic flux, and the area reduction of the naked spots. 

The magnetic field inclination (green in Fig.~\ref{fig:magnetic_evolution}) remained stable from the beginning of the observations. It oscillated around $154\,^{\circ}$ until 23 January at midday. During the rapid decay of the umbral areas and the dissipation of the penumbrae, the inclination became increasingly vertical. Subsequently, during the first stages of naked spots, $\gamma$ oscillated around  $\sim158\,^{\circ}$. The accumulation of new flux led to an increase in horizontal fields along the intensity boundary on average. During the naked-spot stage, $\gamma$ became more vertical on average, diminished, and became more horizontal when the area increased. The same behaviour was observed in our previous case study of the evolution of a pore \citep{garciarivas_pore}.

The magnetic field strength (orange in Fig.~\ref{fig:magnetic_evolution}) was weaker than $B^\mathrm{crit}$=1979~G \citep[found at the boundary of a stable pore;][]{garciarivas_pore} during the whole decaying process. From the beginning of the observations until 23 January midday, \textit{B} oscillated around 1880~G. The oscillations were due to the slow changes that affected the umbral structure. The combination of penumbral fast dissipation and umbral shrinking on 23 January at midday translated into a weakening of $B$. From this moment, $B$ oscillated around a weaker value of 1850~G with fluctuations that were shorter in time. This period corresponds to the first stages of naked spots, while the penumbra completely disappeared. Due to the accumulation of new flux and the consequent increase in the umbral area, $B$ weakened down to its minimum, $\sim$1730~G, after 25 January at 18~UT. Even though there is a tendency to recover the strength of $B$ at the naked-spot boundary and it even reached 1850~G again, the rapid areal changes did not allow for a stable naked-spot stage. In general, $B$ strengthens when the naked-spots area diminishes, and it weakens when the area increases. This again agrees with our previous case study of an evolving pore \citep{garciarivas_pore}.

The vertical field, $B_\mathrm{ver}$ (blue in Fig.~\ref{fig:magnetic_evolution}) behaves very similarly as $B$, although $B_\mathrm{ver}$ shows weaker fluctuations; it is only a scale effect. Similarly to $B$, the magnetic vertical component was weaker in general than $B_\mathrm{ver}^\mathrm{crit}$~=~1821~G, a maximum $B_\mathrm{ver}$ value found by \citep{garciarivas_pore} at the boundary of a stable pore. It punctually reaches the critical value, however. During the first three days, that is, during the slow decay, $B_\mathrm{ver}$ oscillated around 1670~G. Afterwards, the fast penumbral decay began. In the case of $B_\mathrm{ver}$, this induced a strengthening of the vertical field component that led to oscillations around 1690~G during the first stages of naked spots. The highest $B_\mathrm{ver}$ was reached when the northern spot was close to complete dissipation and the southern spot was dark and compact. In these periods, the vertical field reached maximum values. Nonetheless, the accumulation of new flux during the last hours of 25 January caused a sudden drop of $B_\mathrm{ver}$ to below $\sim1550$~G. As for $B$, $B_\mathrm{ver}$ tends to weaken with areal growth and to increase with areal stability or reduction.  

The differences between the temporal trends in $B$ and $B_\mathrm{ver}$ are noteworthy. Both magnetic properties behave in a comparable way, but with a 200~G offset (with $B$ being stronger) during the first observation days. However, this behaviour is only observed when the umbra is surrounded by penumbra. After 24 January at midnight, when the first stages of the naked spot begin and the relative areal changes increase, $B_\mathrm{ver}$ strengthens while $B$ weakens; their offset rises to 250~G on average. The accumulation of new flux again balances the magnetic parameters, but the following decays again increase their offset. We conclude that the presence of a penumbra stabilises the evolution of the magnetic properties at the umbral boundary. After the penumbra has completely disappeared, any change in the spot area translates into significant changes in $B$, $\gamma$, and $B_\mathrm{ver}$. 

The behaviour of $B_\mathrm{ver}$ in the naked spot resembles the evolving pore investigated in \cite{garciarivas_pore}. When the area increases, $B_\mathrm{ver}$ diminishes, and when the area of the spot diminishes, the boundary $B_\mathrm{ver}$ strengthens. In this case study, we do not observe a stable phase nor a critical maximum $B_\mathrm{ver}$, however. During the first days, the sunspot $B_\mathrm{ver}$ oscillates around 1670~G, while the naked-spot flux tube becomes more vertical and punctually reaches $B_\mathrm{ver}=1820$~G. Because of this wide span of values, we explored the evolution of a range of $B_\mathrm{ver}$ thresholds (1600-1850~G) by calculating the areas within the contours, and we compare the evolution of the areas within $I_\mathrm{c,dcon}=0.50I_\mathrm{QS}$ in Fig.~\ref{fig:all_bver}. 

The areas encircled by the intensity contour (red in Fig.~\ref{fig:all_bver}) show a slow but steady decrease during the first three days when the sunspots area decreases from $\mathrm{50~Mm^2}$ to $\mathrm{30~Mm^2}$. An isolated short and sudden decrease is visible on 23 January between approximately 6 and 12\,UT, however, which corresponds to the beginning of the fast penumbral decay in the southern spot. The small area variations within the three-day period are due to the evolution of the umbra-penumbra boundary and to small dark patches embedded in the penumbrae that were only taken into consideration when they were larger than $\mathrm{\sim5~pixel^2}$. This condition changed fast. 

Subsequently, the decay rate of the umbral area suddenly changed during the first hours of 24 January from the previous trend. The umbrae shrink from $\mathrm{30~Mm^2}$ to $\mathrm{10~Mm^2}$ in less than 10 hours. This fast decay coincided with the dissipation of most of the penumbrae in the northern and southern spots, even though penumbral filaments continued to appear and disappear around the northern spot (details of the decay rates are discussed in Sect.~\ref{rates}). The southern spot dimmed, and the umbral area held multiple darker cores. The penumbra completely disappeared during the first hours of 25 January in the northern spot, and the southern pore recovered a compact and darker umbral area. Even though it seemed that the naked spots would completely dissipate, they maintained an area of $\mathrm{\sim10~Mm^2}$ until an accumulation of new flux during the latest hours of 25 January. The areas increased up to $\mathrm{\sim20~Mm^2}$ in approximately 10~hours, but they immediately evolved again, with a final decay at the end of the analysed period.

\begin{figure*}[h!]
  \begin{minipage}[c]{0.75\linewidth}
   \includegraphics[width=\linewidth]{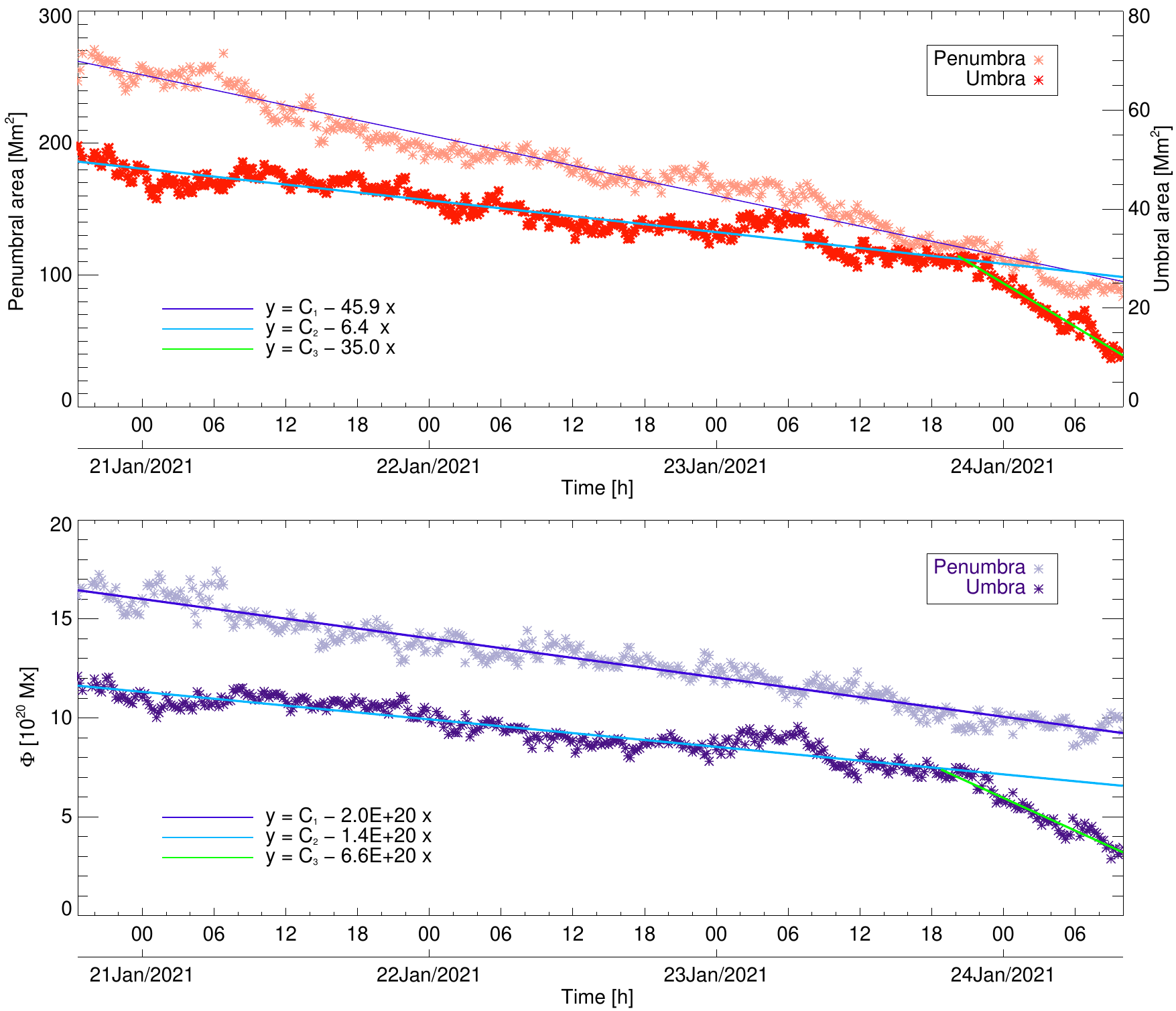}
  \end{minipage}
  \begin{minipage}[c]{0.25\linewidth}
  \vspace{4cm}
    \caption{Temporal evolution of the photometric and magnetic sunspot during the dissipation of the penumbra, which includes the slow and fast umbral decay. Top: Umbral and penumbral areas (red and light red, respectively). Bottom: Umbral and penumbral magnetic flux (dark and light purple, respectively).   }
    \label{fig:decaying_fits}
  \end{minipage}
\end{figure*}

Fig.~\ref{fig:all_bver} does indeed not show a preferred $B_\mathrm{ver}$ on the $I_\mathrm{c,dcon}=0.50I_\mathrm{QS}$ during the whole period. As expected, the slow-decay process of the umbra is closer to the areas encircled by $B_\mathrm{ver}$=1670~G, and the naked spots stages are characterised by stronger vertical fields before the accumulation of new flux. In Fig.~\ref{fig:all_bver}, we visually identify an indication that the fast decay of the area defined by the $B_\mathrm{ver}=1670$~G precedes the fast decay of the area defined by the intensity threshold by $\sim$3.5~hours, as seen between 23 January 18:00~UT and 24 January 6:00~UT. A cross-correlation between the areas encircled by each of the $B_\mathrm{ver}$ thresholds and the area encircled by the intensity contour shows that the areas encircled by $B_\mathrm{ver}$=[1640-1800]\,G have a unique maximum correlation (above 0.975) between 3 and 4 hours, suggesting that the areas that are defined by the vertical field decay 3.5\,h on average before the areas that are defined by the continuum intensity boundary. 

The cross-correlations of the areas that are defined by the intensity threshold and by $B_\mathrm{ver}$\,<\,1640\,G have local maxima in their correlation at a lag of 3.5\,h and also for a lag 0\,h or even for negative lags. The reason might be that the fluctuations in the areas defined by smaller $B_\mathrm{ver}$ are larger since the thresholds are embedded outside the umbral region, and they affect the reliability of the correlation.  

We note that the cross-correlation was made from the beginning of the observations until 25 January at 5\,UT in order to avoid the process of flux accumulation. The variations in the cross-correlation coefficient are low with the time lag (the minimum is 0.96 for positive lags). We recall that the encircled areas are the sum of the two spots that evolve at different paces, and therefore, we combined information that may dilute stronger evidence. In \citet{garciarivas_pore} we did not find any delay between the decay in the areas defined by $B_\mathrm{ver}$ and intensity.

\subsection{Sunspot decay rates}
\label{rates}

In this subsection, we focus on the decay rates of the sunspot, that is, the time at which the penumbra disappears. To do this, we analysed the two components of the sunspot, umbra, and penumbra separately.

Umbral areas are limited by a $I_\mathrm{c,dcon}<0.50I_\mathrm{QS}$ threshold, while penumbral areas are defined by $0.50I_\mathrm{QS}<I_\mathrm{c,dcon}<0.90I_\mathrm{QS}$. The decay rates were investigated from the beginning of the observations until 24 January at 10~UT, when the spots reached their minimum umbral area during the complete dissipation of the penumbra. We note that the disappearance of the penumbra does not imply a lack of spot areas with $0.50<I_\mathrm{QS}<0.90$. That is, in the later stages of the decay, the term penumbral area refers to the brighter regions around the darker umbral areas without penumbral filaments. Figure~\ref{fig:decaying_fits} summarises the temporal evolution of the umbral and penumbral area (red, \textit{A}) and magnetic flux (purple, $\Phi$).

The penumbra (light red, Fig. \ref{fig:decaying_fits}, top) shows a constant linear decay during the whole period at a rate of $\mathrm{45.9~Mm^2 day^{-1}}$. The umbra (red) decays linearly as well, although it shows two stages: (1) a slow decay with a rate of $\mathrm{6.4~Mm^2 day^{-1}}$ from the beginning of the observations to 23 January at $\sim$20~UT, and (2) a fast decay with a rate of $\mathrm{35.0~Mm^2 day^{-1}}$ until the complete disappearance of the penumbrae. The whole sunspot decay is the sum of the decay in the umbra and penumbra. 

The type of decay law depends on the decaying mechanism. For example, the quadratic law is related to the turbulent erosion model, where the dissipation is due to the erosion of the external boundary of the sunspot, and it therefore varies with the length of the boundary. On the other hand, the linear decay law is related to turbulent magnetic diffusion within the whole area of the sunspot \citep[for a wider review see][]{Solanki:2003}. These types of decaying laws have been studied in sunspot groups and in individual sunspots, and the decay rates have been found to be dependent on the sunspot area, the latitude, the solar cycle, or the leading or following sunspot, among others, or to have a lognormal distribution \citep[e.g.][]{1993MartinezPillet_Sunspot_decay_rates, 2020Murakozy_decayRates_umbra, 1997_Petrovay_VanDrielGesztelyi,2008_Hathaway_sunspots_decay}. 

Although statistical analyses are very useful for general rules, they can omit rarer or less noticeable events, such as the detection of different decay rates in one decaying spot. For example, \cite{2016Verma_flowfields_smallAR} studied the case of a small active region that first decayed fast ($\mathrm{23\,Mm^2 day^{-1}}$) and then decayed slowly ($\mathrm{14~Mm^2 day^{-1}}$). Neither the decay processes nor the decay rates agree with the results presented here. A similar result was found by  \cite{2021Qialoning_alphas_decay} in one of the eight $\alpha$-sunspots they analysed, while seven sunspots decayed slowly, followed by a fast linear decay, like in our study. The sunspot decay we studied first decayed slowly and then fast only due to a change in the umbral decay, but \cite{2021Qialoning_alphas_decay} reported diverse scenarios (constant umbral decay and two-stage penumbral decay; two-stage umbral decay and two-stage penumbral decay). Even their reported decay rates (slow umbral decay: $\mathrm{3.3\,Mm^2 day^{-1}}$, fast umbral decay: $\mathrm{29\,Mm^2 day^{-1}}$, and fast penumbral decay: $\mathrm{38\,Mm^2 day^{-1}}$) are lower than ours, perhaps due to the difference in the sunspot areas. 

The ratio of the umbra-to-penumbra area (U/P), the umbra-to-total area, or the penumbra-to-umbra area were used in the literature to describe sunspot decay that take the evolution of their components into account. It has been found to be constant \citep[][e.g.]{1993MartinezPillet_Sunspot_decay_rates} and variable, depending on the sunspot decay rate. \cite{2018_Carrasco_UP_ratio_in_MauderMinimum} found higher U/P values during faster sunspot decays, for instance, while \cite{2003Champan_decay_sunspots} found higher U/P ratios with slower sunspot decays. The sunspot we studied has a variable U/P ratio that varies with the velocity of the umbral decay. During the slow umbral decay, the U/P ratio oscillates around 0.21, while during the fast umbral decay, it reaches its minimum (0.13). However, the sunspot decay rate remains constant, and therefore, the U/P ratio is lower during a constant sunspot decay. Similar results were found by \cite{2021Qialoning_alphas_decay} in certain sunspots, where the significant drop of the U/P ratio was connected to sudden drops of the umbral area and not to sudden changes in the evolution of the whole sunspot.

The flux trend is parallel to the decaying areas: The penumbral flux (light purple in Fig. \ref{fig:decaying_fits}, bottom) decays linearly during the whole period we studied at a rate of  $\mathrm{2\,\times\,10^{20}\,Mx\,day^{-1}}$, and the umbral flux (dark purple) decays in two stages, $\mathrm{1.4\,\times\,10^{20}\,Mx\,day^{-1}}$ at the beginning, and $\mathrm{6.6\,\times\,10^{20}\,Mx\,day^{-1}}$ during the fast decay. Compared to \cite{2021Qialoning_alphas_decay}, our umbral flux decay rates are twice higher, but their penumbral flux decreases faster. Our slow-decay rate is similar to the rate derived by \cite{2007_Deng_decaying_sunspot} ($\mathrm{2.8\,\times\,10^{20}\,Mx\,day^{-1}}$), while our fast-decay rate is closer to the decay rate of a naked spot simulated by \cite{2015Rempel_decaySunspot_nakedspot}. Overall, \cite{Benko_etal2018} reported a similar flux disappearance to the values we find during the slow umbral and penumbral decay. 


\section{Discussion and conclusions}\label{sec:conclusions}

We presented the analysis of a decaying sunspot that turned into naked spots while transiting from the eastern to the western solar limb. The aim of this work was to analyse the evolution of the magnetic properties on the UP boundary during the decay phase. Moreover, this sunspot enabled us to compare the differences between datasets that were previously used in the study of UP boundaries, that is, SP maps inferred with SIR and MERLIN, regular HMI maps, and $\mathrm{HMI_{dcon}}$ maps. Not only can we compare the dissimilarities among the derived properties inside the spot, but we can concentrate on the divergences at the UP boundary in particular.

Due to the choice of a unique continuum intensity threshold for our study, we find that UP boundaries are better defined by the $0.50\,I_\mathrm{QS}$ threshold than the naked spots. As reported in \cite{garciarivas_pore}, pores are more precisely defined by a $0.55\,I_\mathrm{QS}$ threshold, especially when the pore structure is compact.  

From a comparison of the intensity thresholds defined individually for each dataset at $0.50\,I_\mathrm{QS}$ (Fig. \ref{fig:maps_all_contours}), we see that HMI outlines only the larger structures without fine details and fails to characterise the smaller structures. During the naked-spot stage, the HMI $0.50\,I_\mathrm{QS}$ threshold falls well inside the spot. On the other hand, the $\mathrm{HMI_{dcon}}$ and SP $0.50\,I_\mathrm{QS}$ contours are more similar, even though the SP contours have more finely detailed structures. The lower spatial resolution of $\mathrm{HMI_{dcon}}$ translates into the apparent fusion of structures that SP considers as separated structures. As proposed in \cite{garciarivas_pore} and proven in this study, $\mathrm{HMI_{dcon}}$ and SP are a more accurate source for studying the UP and pore boundaries. 

From a general comparison of the spot properties shown in Fig.~\ref{fig:scatter_plots} and from the detailed comparison of the physical properties found on the UP boundary shown in Fig.~\ref{fig:050HMI_inALL}, we find the best agreement between the studied datasets in the $B_\mathrm{ver}$ parameter, while the continuum intensity lacks a good one-to-one correspondence like this and shows larger scatter. This implies that a fixed $B_\mathrm{ver}$ threshold would define the areas in all the datasets more accurately than a fixed $I_\mathrm{c}$ threshold.

The UP boundary defined in the HMI$_\mathrm{dcon}$ dataset at $I_\mathrm{c}=0.50I_\mathrm{QS}$ is characterised in SP maps by $I_\mathrm{c}=0.54I_\mathrm{QS}$ (8\% brighter). This intensity value fluctuates from one map to the next, but it is not dependent on the spot structure, size, or location. On the other hand, the $I_\mathrm{c}$ value found in HMI maps brightens with a smaller spot size, that is, when more scattered light from the brighter granulation contaminates the darker structures. The intensities found at the boundaries in HMI maps increase from $I_\mathrm{c}=0.55I_\mathrm{QS}$, 10\% brighter, to $I_\mathrm{c}=0.58I_\mathrm{QS}$, 16\%brighter (see Fig.~\ref{fig:050HMI_inALL}).

Along with the differences of $I_\mathrm{c}$ between the analysed datasets, we also found differences in the $B_\mathrm{ver}$, values as shown in Fig.~\ref{fig:050HMI_inALL}. These differences explain the origin of the discrepancies in the $B_\mathrm{ver}$ values found at the umbral boundaries in previous studies. 

The mean $B_\mathrm{ver}$ at the UP boundary in the SP maps is 5\,\% weaker on average than the value obtained in $\mathrm{HMI_{dcon}}$ maps. The fact that $\mathrm{HMI_{dcon}}$ maps provide the strongest $B_\mathrm{ver}$ values implies that the maximum vertical field found in the stable stage of an evolving pore ($B_\mathrm{ver}^\mathrm{crit}$=1821\,G at $0.50\,I_\mathrm{QS}$) by \cite{garciarivas_pore} is not comparable to the $B_\mathrm{ver}^\mathrm{crit}$ value obtained from a statistical analysis of stable sunspots in SP maps (1867 $\pm$ 18\,G)  by \cite{Jurcak_etal2018}.

On the other hand, the difference of $B_\mathrm{ver}$ at the UP boundary between $\mathrm{HMI_{dcon}}$ and HMI maps largely depends on the spot size (it increases with smaller areas), and it varies from 8\,\% (sunspot) to 15\,\% (naked spots). The disagreement between the $B_\mathrm{ver}^\mathrm{crit}$ obtained by \cite{Schmassmann_etal2018} in a stable sunspot in HMI maps with respect to the one obtained by \cite{garciarivas_pore} (1695\,G and 1821\,G at $0.50\,I_\mathrm{QS}$, respectively) is 6\,\%, which agrees with the dissimilarities between the sunspot stage in this study. Complementary, the $B_\mathrm{ver}^\mathrm{crit}$ obtained by \cite{Schmassmann_etal2018} is 9\,\% weaker than the value obtained by \cite{Jurcak_etal2018}, which again agrees with the difference between $B_\mathrm{ver}$ at the UP boundary of HMI with respect to SP maps found during the sunspot stage in this study (5\,\%).

As reported in previous studies during the evolution of pores \citep{garciarivas_pore}, an increase in the naked-spots area translates into drops of the magnetic fields at the boundary. On the other hand, non-variable or decreasing areas yield increasing $B_\mathrm{ver}$ values at the $I_\mathrm{c}$ threshold during non-stable stages. The maximum $B_\mathrm{ver}$ values are achieved in the naked-spot stage, even though it never reaches the $B_\mathrm{ver}^\mathrm{stable}$ found in previous studies, nor is a stable period in terms of constant area observed.

An interesting result is that the areas encircled by $B_\mathrm{ver}$ thresholds between 1640\,G and 1800\,G seem to decay $\sim$3.5\,h before the areas encircled by the intensity contour (Fig. \ref{fig:all_bver}). This result indicates that the magnetic field may decay before the photometric structure. However, the correlation coefficient that indicates the earlier decay of the magnetic field depends only weakly on the time lag. 

The area and magnetic flux decay were studied separately in the umbra and in the penumbra. The photometric spot and the $\Phi$ decay linearly, which suggests magnetic diffusion in the whole area of the sunspot \citep[e.g.][]{Solanki:2003}. While the penumbra decays at a constant rate, the umbra first decays slowly, which is followed by a sudden fast decay in the later stages of the penumbral dissipation. This behaviour is not consistent with reported constant linear or quadratic decay rates \citep[e.g.][]{Benko_etal2018, 2007_Deng_decaying_sunspot, 1997_Petrovay_VanDrielGesztelyi}. Nonetheless, decaying processes divided into slow and fast stages have been observed as well \citep[e.g.][]{2021Qialoning_alphas_decay,2016Verma_flowfields_smallAR, garciarivas_pore}, even though the order (slow/fast or fast/slow) seems to depend on the case study. 

\begin{acknowledgements}
    We would like to thank A. Norton for providing us with deconvolved full-disc HMI vector field maps. This work was supported by the Czech-German common grant, funded by the Czech Science Foundation under the project 23-07633K and by the Deutsche Forschungsgemeinschaft under the project BE 5771/3-1 (eBer-23-13412), and the institutional support ASU:67985815 of the Czech Academy of Sciences. The HMI data are courtesy of NASA/SDO and the HMI science teams. Hinode is a Japanese mission developed and launched by ISAS/JAXA, collaborating with NAOJ as a domestic partner, NASA and STFC (UK) as international partners. Scientific operation of the Hinode mission is conducted by the Hinode science team organized at ISAS/JAXA. This team mainly consists of scientists from institutes in the partner countries. Support for the post-launch operation is provided by JAXA and NAOJ(Japan), STFC (U.K.), NASA, ESA, and NSC (Norway).
\end{acknowledgements}

\bibliographystyle{bibtex/aa} 
\bibliography{bibtex/biblio} 

\begin{thebibliography}{52}
\expandafter\ifx\csname natexlab\endcsname\relax\def\natexlab#1{#1}\fi

\bibitem[{{Andretta} {et~al.}(2021){Andretta}, {Bemporad}, {De Leo}, {Jerse}, {Landini}, {Mierla}, {Naletto}, {Romoli}, {Sasso}, {Slemer}, {Spadaro}, {Susino}, {Talpeanu}, {Telloni}, {Teriaca}, {Uslenghi}, {Antonucci}, {Auch{\`e}re}, {Berghmans}, {Berlicki}, {Capobianco}, {Capuano}, {Casini}, {Casti}, {Chioetto}, {Da Deppo}, {Fabi}, {Fineschi}, {Frassati}, {Frassetto}, {Giordano}, {Grimani}, {Heinzel}, {Liberatore}, {Magli}, {Massone}, {Messerotti}, {Moses}, {Nicolini}, {Pancrazzi}, {Pelizzo}, {Romano}, {Sch{\"u}hle}, {Stangalini}, {Straus}, {Volpicelli}, {Zangrilli}, {Zuppella}, {Abbo}, {Aznar Cuadrado}, {Bruno}, {Ciaravella}, {D'Amicis}, {Lamy}, {Lanzafame}, {Malvezzi}, {Nicolosi}, {Nistic{\`o}}, {Peter}, {Plainaki}, {Poletto}, {Reale}, {Solanki}, {Strachan}, {Tondello}, {Tsinganos}, {Velli}, {Ventura}, {Vial}, {Woch}, \& {Zimbardo}}]{NOAA12797_Metis}
{Andretta}, V., {Bemporad}, A., {De Leo}, Y., {et~al.} 2021, \aap, 656, L14

\bibitem[{{Antonucci} {et~al.}(2020){Antonucci}, {Romoli}, {Andretta}, {Fineschi}, {Heinzel}, {Moses}, {Naletto}, {Nicolini}, {Spadaro}, {Teriaca}, {Berlicki}, {Capobianco}, {Crescenzio}, {Da Deppo}, {Focardi}, {Frassetto}, {Heerlein}, {Landini}, {Magli}, {Marco Malvezzi}, {Massone}, {Melich}, {Nicolosi}, {Noci}, {Pancrazzi}, {Pelizzo}, {Poletto}, {Sasso}, {Sch{\"u}hle}, {Solanki}, {Strachan}, {Susino}, {Tondello}, {Uslenghi}, {Woch}, {Abbo}, {Bemporad}, {Casti}, {Dolei}, {Grimani}, {Messerotti}, {Ricci}, {Straus}, {Telloni}, {Zuppella}, {Auch{\`e}re}, {Bruno}, {Ciaravella}, {Corso}, {Alvarez Copano}, {Aznar Cuadrado}, {D'Amicis}, {Enge}, {Gravina}, {Jej{\v{c}}i{\v{c}}}, {Lamy}, {Lanzafame}, {Meierdierks}, {Papagiannaki}, {Peter}, {Fernandez Rico}, {Giday Sertsu}, {Staub}, {Tsinganos}, {Velli}, {Ventura}, {Verroi}, {Vial}, {Vives}, {Volpicelli}, {Werner}, {Zerr}, {Negri}, {Castronuovo}, {Gabrielli}, {Bertacin}, {Carpentiero}, {Natalucci}, {Marliani}, {Cesa}, {Laget}, {Morea}, {Pieraccini}, {Radaelli},
  {Sandri}, {Sarra}, {Cesare}, {Del Forno}, {Massa}, {Montabone}, {Mottini}, {Quattropani}, {Schillaci}, {Boccardo}, {Brando}, {Pandi}, {Baietto}, {Bertone}, {Alvarez-Herrero}, {Garc{\'\i}a Parejo}, {Cebollero}, {Amoruso}, \& {Centonze}}]{2020A_Antonucci_Metis}
{Antonucci}, E., {Romoli}, M., {Andretta}, V., {et~al.} 2020, \aap, 642, A10

\bibitem[{{Balthasar}(2018)}]{Balthasar:2018}
{Balthasar}, H. 2018, \solphys, 293, 120

\bibitem[{{Beck} \& {Chapman}(1993)}]{1993_Beck}
{Beck}, J.~G. \& {Chapman}, G.~A. 1993, \solphys, 146, 49

\bibitem[{{Benko} {et~al.}(2018){Benko}, {Gonz{\'a}lez Manrique}, {Balthasar}, {G{\"o}m{\"o}ry}, {Kuckein}, \& {Jur{\v{c}}{\'a}k}}]{Benko_etal2018}
{Benko}, M., {Gonz{\'a}lez Manrique}, S.~J., {Balthasar}, H., {et~al.} 2018, \aap, 620, A191

\bibitem[{{Borrero} {et~al.}(2011){Borrero}, {Tomczyk}, {Kubo}, {Socas-Navarro}, {Schou}, {Couvidat}, \& {Bogart}}]{VFISV_Borrero2011}
{Borrero}, J.~M., {Tomczyk}, S., {Kubo}, M., {et~al.} 2011, \solphys, 273, 267

\bibitem[{{Cabrera Solana} {et~al.}(2005){Cabrera Solana}, {Bellot Rubio}, \& {del Toro Iniesta}}]{CabreraSolana:2005}
{Cabrera Solana}, D., {Bellot Rubio}, L.~R., \& {del Toro Iniesta}, J.~C. 2005, \aap, 439, 687

\bibitem[{{Campos Rozo} {et~al.}(2023){Campos Rozo}, {Vargas Dom{\'\i}nguez}, {Utz}, {Veronig}, \& {Hanslmeier}}]{2023_Poros_JoseIvan}
{Campos Rozo}, J.~I., {Vargas Dom{\'\i}nguez}, S., {Utz}, D., {Veronig}, A.~M., \& {Hanslmeier}, A. 2023, \aap, 674, A91

\bibitem[{{Carrasco} {et~al.}(2018){Carrasco}, {Garc{\'\i}a-Romero}, {Vaquero}, {Rodr{\'\i}guez}, {Foukal}, {Gallego}, \& {Lef{\`e}vre}}]{2018_Carrasco_UP_ratio_in_MauderMinimum}
{Carrasco}, V.~M.~S., {Garc{\'\i}a-Romero}, J.~M., {Vaquero}, J.~M., {et~al.} 2018, \apj, 865, 88

\bibitem[{{Centeno} {et~al.}(2009){Centeno}, {Lites}, {de Wijn}, \& {Elmore}}]{2009_Centeno_noregslitscan_SP}
{Centeno}, R., {Lites}, B., {de Wijn}, A.~G., \& {Elmore}, D. 2009, in Astronomical Society of the Pacific Conference Series, Vol. 415, The Second Hinode Science Meeting: Beyond Discovery-Toward Understanding, ed. B.~{Lites}, M.~{Cheung}, T.~{Magara}, J.~{Mariska}, \& K.~{Reeves}, 323

\bibitem[{{Centeno} {et~al.}(2014){Centeno}, {Schou}, {Hayashi}, {Norton}, {Hoeksema}, {Liu}, {Leka}, \& {Barnes}}]{VFISV_Centeno2014}
{Centeno}, R., {Schou}, J., {Hayashi}, K., {et~al.} 2014, \solphys, 289, 3531

\bibitem[{{Chandrasekhar}(1961)}]{Chandrasekhar1961}
{Chandrasekhar}, S. 1961, {Hydrodynamic and hydromagnetic stability}

\bibitem[{{Chapman} {et~al.}(2003){Chapman}, {Dobias}, {Preminger}, \& {Walton}}]{2003Champan_decay_sunspots}
{Chapman}, G.~A., {Dobias}, J.~J., {Preminger}, D.~G., \& {Walton}, S.~R. 2003, \grl, 30, 1178

\bibitem[{{Criscuoli} {et~al.}(2017){Criscuoli}, {Norton}, \& {Whitney}}]{HMI_dcon_Norton}
{Criscuoli}, S., {Norton}, A., \& {Whitney}, T. 2017, \apj, 847, 93

\bibitem[{{Deng} {et~al.}(2007){Deng}, {Choudhary}, {Tritschler}, {Denker}, {Liu}, \& {Wang}}]{2007_Deng_decaying_sunspot}
{Deng}, N., {Choudhary}, D.~P., {Tritschler}, A., {et~al.} 2007, \apj, 671, 1013

\bibitem[{{Garc{\'\i}a-Rivas} {et~al.}(2021){Garc{\'\i}a-Rivas}, {Jur{\v{c}}{\'a}k}, \& {Bello Gonz{\'a}lez}}]{garciarivas_pore}
{Garc{\'\i}a-Rivas}, M., {Jur{\v{c}}{\'a}k}, J., \& {Bello Gonz{\'a}lez}, N. 2021, \aap, 649, A129

\bibitem[{{Gough} \& {Tayler}(1966)}]{Gough:1966}
{Gough}, D.~O. \& {Tayler}, R.~J. 1966, \mnras, 133, 85

\bibitem[{{Grossmann-Doerth} \& {Schmidt}(1981)}]{1981_Grossmann-Doerth}
{Grossmann-Doerth}, U. \& {Schmidt}, W. 1981, \aap, 95, 366

\bibitem[{{Hathaway} \& {Choudhary}(2008)}]{2008_Hathaway_sunspots_decay}
{Hathaway}, D.~H. \& {Choudhary}, D.~P. 2008, \solphys, 250, 269

\bibitem[{{Hoeksema} {et~al.}(2014){Hoeksema}, {Liu}, {Hayashi}, {Sun}, {Schou}, {Couvidat}, {Norton}, {Bobra}, {Centeno}, {Leka}, {Barnes}, \& {Turmon}}]{HMI_Pipeline_Hoeksema2014}
{Hoeksema}, J.~T., {Liu}, Y., {Hayashi}, K., {et~al.} 2014, \solphys, 289, 3483

\bibitem[{{Ichimoto} {et~al.}(2008){Ichimoto}, {Lites}, {Elmore}, {Suematsu}, {Tsuneta}, {Katsukawa}, {Shimizu}, {Shine}, {Tarbell}, {Title}, {Kiyohara}, {Shinoda}, {Card}, {Lecinski}, {Streander}, {Nakagiri}, {Miyashita}, {Noguchi}, {Hoffmann}, \& {Cruz}}]{Ichimoto2008}
{Ichimoto}, K., {Lites}, B., {Elmore}, D., {et~al.} 2008, \solphys, 249, 233

\bibitem[{{Jur{\v{c}}{\'a}k}(2011)}]{Jurcak2011}
{Jur{\v{c}}{\'a}k}, J. 2011, \aap, 531, A118

\bibitem[{{Jur{\v{c}}{\'a}k} {et~al.}(2015){Jur{\v{c}}{\'a}k}, {Bello Gonz{\'a}lez}, {Schlichenmaier}, \& {Rezaei}}]{Jurcak_etal2015}
{Jur{\v{c}}{\'a}k}, J., {Bello Gonz{\'a}lez}, N., {Schlichenmaier}, R., \& {Rezaei}, R. 2015, \aap, 580, L1

\bibitem[{{Jur{\v{c}}{\'a}k} {et~al.}(2017){Jur{\v{c}}{\'a}k}, {Bello Gonz{\'a}lez}, {Schlichenmaier}, \& {Rezaei}}]{Jurcak_etal2017}
{Jur{\v{c}}{\'a}k}, J., {Bello Gonz{\'a}lez}, N., {Schlichenmaier}, R., \& {Rezaei}, R. 2017, \aap, 597, A60

\bibitem[{{Jur{\v{c}}{\'a}k} {et~al.}(2018){Jur{\v{c}}{\'a}k}, {Rezaei}, {Gonz{\'a}lez}, {Schlichenmaier}, \& {Vomlel}}]{Jurcak_etal2018}
{Jur{\v{c}}{\'a}k}, J., {Rezaei}, R., {Gonz{\'a}lez}, N.~B., {Schlichenmaier}, R., \& {Vomlel}, J. 2018, \aap, 611, L4

\bibitem[{{Kosugi} {et~al.}(2007){Kosugi}, {Matsuzaki}, {Sakao}, {Shimizu}, {Sone}, {Tachikawa}, {Hashimoto}, {Minesugi}, {Ohnishi}, {Yamada}, {Tsuneta}, {Hara}, {Ichimoto}, {Suematsu}, {Shimojo}, {Watanabe}, {Shimada}, {Davis}, {Hill}, {Owens}, {Title}, {Culhane}, {Harra}, {Doschek}, \& {Golub}}]{Hinode_Kosugi2007}
{Kosugi}, T., {Matsuzaki}, K., {Sakao}, T., {et~al.} 2007, \solphys, 243, 3

\bibitem[{{Leka} {et~al.}(2009){Leka}, {Barnes}, {Crouch}, {Metcalf}, {Gary}, {Jing}, \& {Liu}}]{disambiguation_leka}
{Leka}, K.~D., {Barnes}, G., {Crouch}, A.~D., {et~al.} 2009, \solphys, 260, 83

\bibitem[{{Li} {et~al.}(2022){Li}, {Zhang}, {Yan}, {Norton}, {Wang}, {Yang}, {Xue}, \& {Kong}}]{2022_Li}
{Li}, Q., {Zhang}, L., {Yan}, X., {et~al.} 2022, \apj, 936, 37

\bibitem[{{Li} {et~al.}(2021){Li}, {Zhang}, {Yan}, {Wang}, {Kong}, {Yang}, \& {Xue}}]{2021Qialoning_alphas_decay}
{Li}, Q., {Zhang}, L., {Yan}, X., {et~al.} 2021, \apj, 913, 147

\bibitem[{{Lindner} {et~al.}(2020){Lindner}, {Schlichenmaier}, \& {Bello Gonz{\'a}lez}}]{Lindner:2020}
{Lindner}, P., {Schlichenmaier}, R., \& {Bello Gonz{\'a}lez}, N. 2020, \aap, 638, A25

\bibitem[{{Lites} {et~al.}(2007){Lites}, {Casini}, {Garcia}, \& {Socas-Navarro}}]{2007_Lites_MERLIN}
{Lites}, B., {Casini}, R., {Garcia}, J., \& {Socas-Navarro}, H. 2007, \memsai, 78, 148

\bibitem[{{L{\"o}ptien} {et~al.}(2018){L{\"o}ptien}, {Lagg}, {van Noort}, \& {Solanki}}]{Loptien:2018}
{L{\"o}ptien}, B., {Lagg}, A., {van Noort}, M., \& {Solanki}, S.~K. 2018, \aap, 619, A42

\bibitem[{{L{\"o}ptien} {et~al.}(2020){L{\"o}ptien}, {Lagg}, {van Noort}, \& {Solanki}}]{Loptien2020_NoUniversalConnection}
{L{\"o}ptien}, B., {Lagg}, A., {van Noort}, M., \& {Solanki}, S.~K. 2020, \aap, 639, A106

\bibitem[{{Mart{\'\i}nez Pillet}(2002)}]{2002MartinezPillet_decaySunspotMMF}
{Mart{\'\i}nez Pillet}, V. 2002, Astronomische Nachrichten, 323, 342

\bibitem[{{Martinez Pillet} {et~al.}(1993){Martinez Pillet}, {Moreno-Insertis}, \& {Vazquez}}]{1993MartinezPillet_Sunspot_decay_rates}
{Martinez Pillet}, V., {Moreno-Insertis}, F., \& {Vazquez}, M. 1993, \aap, 274, 521

\bibitem[{{Mathew} {et~al.}(2007){Mathew}, {Mart{\'\i}nez Pillet}, {Solanki}, \& {Krivova}}]{2007_Mathew}
{Mathew}, S.~K., {Mart{\'\i}nez Pillet}, V., {Solanki}, S.~K., \& {Krivova}, N.~A. 2007, \aap, 465, 291

\bibitem[{{Metcalf}(1994)}]{disambiguation_metcalf}
{Metcalf}, T.~R. 1994, \solphys, 155, 235

\bibitem[{{M{\"u}ller} {et~al.}(2020){M{\"u}ller}, {St. Cyr}, {Zouganelis}, {Gilbert}, {Marsden}, {Nieves-Chinchilla}, {Antonucci}, {Auch{\`e}re}, {Berghmans}, {Horbury}, {Howard}, {Krucker}, {Maksimovic}, {Owen}, {Rochus}, {Rodriguez-Pacheco}, {Romoli}, {Solanki}, {Bruno}, {Carlsson}, {Fludra}, {Harra}, {Hassler}, {Livi}, {Louarn}, {Peter}, {Sch{\"u}hle}, {Teriaca}, {del Toro Iniesta}, {Wimmer-Schweingruber}, {Marsch}, {Velli}, {De Groof}, {Walsh}, \& {Williams}}]{2020A_Muller_SolarOrbiter}
{M{\"u}ller}, D., {St. Cyr}, O.~C., {Zouganelis}, I., {et~al.} 2020, \aap, 642, A1

\bibitem[{{Murak{\"o}zy}(2020)}]{2020Murakozy_decayRates_umbra}
{Murak{\"o}zy}, J. 2020, \apj, 892, 107

\bibitem[{{Pesnell} {et~al.}(2012){Pesnell}, {Thompson}, \& {Chamberlin}}]{SDO_Pesnell2012}
{Pesnell}, W.~D., {Thompson}, B.~J., \& {Chamberlin}, P.~C. 2012, \solphys, 275, 3

\bibitem[{{Petrovay} \& {van Driel-Gesztelyi}(1997)}]{1997_Petrovay_VanDrielGesztelyi}
{Petrovay}, K. \& {van Driel-Gesztelyi}, L. 1997, \solphys, 176, 249

\bibitem[{{Pettauer} \& {Brandt}(1997)}]{1997_Pettauer_cumulativeHist}
{Pettauer}, T. \& {Brandt}, P.~N. 1997, \solphys, 175, 197

\bibitem[{{Rempel}(2015)}]{2015Rempel_decaySunspot_nakedspot}
{Rempel}, M. 2015, \apj, 814, 125

\bibitem[{{Romano} {et~al.}(2020){Romano}, {Murabito}, {Guglielmino}, {Zuccarello}, \& {Falco}}]{2020Romano_restoringPenumbra}
{Romano}, P., {Murabito}, M., {Guglielmino}, S.~L., {Zuccarello}, F., \& {Falco}, M. 2020, \apj, 899, 129

\bibitem[{{Ruiz Cobo} \& {del Toro Iniesta}(1992)}]{SIR_RuizCobo1992}
{Ruiz Cobo}, B. \& {del Toro Iniesta}, J.~C. 1992, \apj, 398, 375

\bibitem[{{Schmassmann} {et~al.}(2021){Schmassmann}, {Rempel}, {Bello Gonz{\'a}lez}, {Schlichenmaier}, \& {Jur{\v{c}}{\'a}k}}]{Schmassmann_mhd}
{Schmassmann}, M., {Rempel}, M., {Bello Gonz{\'a}lez}, N., {Schlichenmaier}, R., \& {Jur{\v{c}}{\'a}k}, J. 2021, \aap, 656, A92

\bibitem[{{Schmassmann} {et~al.}(2018){Schmassmann}, {Schlichenmaier}, \& {Bello Gonz{\'a}lez}}]{Schmassmann_etal2018}
{Schmassmann}, M., {Schlichenmaier}, R., \& {Bello Gonz{\'a}lez}, N. 2018, \aap, 620, A104

\bibitem[{{Schou} {et~al.}(2012){Schou}, {Scherrer}, {Bush}, {Wachter}, {Couvidat}, {Rabello-Soares}, {Bogart}, {Hoeksema}, {Liu}, {Duvall}, {Akin}, {Allard}, {Miles}, {Rairden}, {Shine}, {Tarbell}, {Title}, {Wolfson}, {Elmore}, {Norton}, \& {Tomczyk}}]{HMI_Schou2012}
{Schou}, J., {Scherrer}, P.~H., {Bush}, R.~I., {et~al.} 2012, \solphys, 275, 229

\bibitem[{{Solanki}(2003)}]{Solanki:2003}
{Solanki}, S.~K. 2003, \aapr, 11, 153

\bibitem[{{Strecker} {et~al.}(2021){Strecker}, {Schmidt}, {Schlichenmaier}, \& {Rempel}}]{Strecker+2021}
{Strecker}, H., {Schmidt}, W., {Schlichenmaier}, R., \& {Rempel}, M. 2021, \aap, 649, A123

\bibitem[{{Tsuneta} {et~al.}(2008){Tsuneta}, {Ichimoto}, {Katsukawa}, {Nagata}, {Otsubo}, {Shimizu}, {Suematsu}, {Nakagiri}, {Noguchi}, {Tarbell}, {Title}, {Shine}, {Rosenberg}, {Hoffmann}, {Jurcevich}, {Kushner}, {Levay}, {Lites}, {Elmore}, {Matsushita}, {Kawaguchi}, {Saito}, {Mikami}, {Hill}, \& {Owens}}]{SOT_Tsuneta2008}
{Tsuneta}, S., {Ichimoto}, K., {Katsukawa}, Y., {et~al.} 2008, \solphys, 249, 167

\bibitem[{{Verma} {et~al.}(2016){Verma}, {Denker}, {Balthasar}, {Kuckein}, {Gonz{\'a}lez Manrique}, {Sobotka}, {Bello Gonz{\'a}lez}, {Hoch}, {Diercke}, {Kummerow}, {Berkefeld}, {Collados}, {Feller}, {Hofmann}, {Kneer}, {Lagg}, {L{\"o}hner-B{\"o}ttcher}, {Nicklas}, {Pastor Yabar}, {Schlichenmaier}, {Schmidt}, {Schmidt}, {Schubert}, {Sigwarth}, {Solanki}, {Soltau}, {Staude}, {Strassmeier}, {Volkmer}, {von der L{\"u}he}, \& {Waldmann}}]{2016Verma_flowfields_smallAR}
{Verma}, M., {Denker}, C., {Balthasar}, H., {et~al.} 2016, \aap, 596, A3

\end{thebibliography}

\end{document}